\documentclass[nofootinbib,preprint,preprintnumbers,a4paper,11pt]{revtex4}
\usepackage{textcomp}
\usepackage{epsfig}
\usepackage{footnote}
\usepackage{ulem}
\usepackage{color}
\usepackage{array}
\usepackage{amssymb}
\usepackage{amsmath}
\usepackage{graphicx}
\usepackage{subfigure}
\usepackage{longtable}
\usepackage{verbatim}
\usepackage{amsfonts}
\usepackage{hyperref}
\usepackage{multirow}
\usepackage{cancel}

\begin{document}

\title{Isospin sum rules for the nonleptonic $B$ decays}

\author{Di Wang$^{1}$}\email{wangdi@hunnu.edu.cn}

\address{%
$^1$Department of Physics, Hunan Normal University, Changsha 410081, China
}

\begin{abstract}
Isospin symmetry, as the most precise flavor symmetry, can be used to extract information about hadronic dynamics.
The effective Hamiltonian for bottom quark weak decay is zero under the isospin lowering operators $I_-^n$, which allows us to generate isospin sum rules through several master formulas.
In this work, we derive the master formulas of isospin sum rules for the two- and three-body non-leptonic decays of $B$ mesons.
Numerous new isospin sum rules, especially those involving three or more decay channels, are proposed.
The isospin sum rules can be used to test isospin symmetry and provide hints the isospin partners of exotic hadrons in $B$ decays.
It is found ten percent isospin breaking only in decay modes involving two vector mesons, indicating the complex dynamics of vector mesons.
Besides, the isospin analysis suggests the charm tetraquark resonances might be observed in the $ B^- \to J/\Psi \pi^-\overline K^0$, $\overline B^0 \to J/\Psi \overline K^0\phi$, and $\overline B^0\to D^0\overline D^0\overline K^0$ modes.

\end{abstract}

\maketitle

\section{Introduction}

$B$ meson weak decay provides an ideal laboratory to study weak and strong interactions and search for new physics beyond the Standard Model.
Flavor symmetry is a powerful tool to analyze heavy hadron weak decays, which has been widely applied to $B$ meson decays \cite{Savage:1989ub,Gronau:1990ka,Qin:2022nof,Gavrilova:2022hbx,Hassan:2022ucn,Wang:2020gmn,He:2018joe,He:2018php,Ligeti:2015yma,Grossman:2013lya,Gronau:2000zy,Jung:2009pb}.
Flavor symmetry leads to sum rules for several decay amplitudes.
Isospin symmetry is the most precise flavor symmetry.
Isospin breaking is naively expected as $\delta_I \simeq (m_u-m_d)/\Lambda_{\rm QCD}\sim 1\%$.
Isospin sum rules could provide valuable insights into the hadronic dynamics of $B$ meson decays.
For example, the isospin sum rule for the $B\to\pi\pi$ system is critical in the Gronau-London method \cite{Gronau:1990ka} for determining the Cabibbo-Kobayashi-Maskawa (CKM) quark mixing parameter $\alpha\equiv Arg[V_{td}V^*_{tb}/V_{ud}V^*_{ub}]$.

Isospin sum rules are typically derived by observing the decay amplitudes expressed by the Wigner-Eckhart invariants \cite{Eckart30,Wigner59}.
For example, the isospin sum rule for the $\overline B\to \pi\pi$ system is obtained from the isospin decompositions of the $B^-\to \pi^0\pi^-$, $\overline B^0\to \pi^+\pi^-$ and $\overline B^0\to \pi^0\pi^0$ modes.
In Refs.~\cite{Wang:2023pnb,Wang:2023don}, we propose a simple approach to generate isospin sum rules for the heavy hadron decays without relying on the isospin amplitudes.
The effective Hamiltonian for heavy quark weak decay is zero under isospin lowering operators $I_-^n$. This fact allows us to generate isospin sum rules using several master formulas.
In this work, we apply this approach to the two- and three-body decays of $B$ mesons.
The master formulas for generating isospin sum rules are derived.
Many new isospin sum rules, especially those involving three or more decay channels, are found by these master formulas.
The isospin sum rules could serve as a test of the isospin symmetry and provide hints for the isospin partners of exotic hadrons through $B$ meson decays.

The rest of this paper is structured as follows.
The theoretical framework and the master formulas for isospin sum rules in $B$ decays are presented in Sec. \ref{IB}.
The phenomenological analysis of the isospin sum rules is discussed in Sec. \ref{PA}.
And Sec. \ref{SUM} is a brief summary.
The isospin sum rules for the two- and three-body $B$ meson decays are listed in Appendix \ref{rules}.

\section{Generation of isospin sum rule for $B$ decay}\label{IB}

Taking the $\overline B\to DM$ decays (where $D$ is a charm meson and $M$ is a light meson) as examples, we demonstrate the basic idea of generating isospin sum rules by $I_-^n$ as follows.
The effective Hamiltonian of $b\to c\overline u q$ transition is given by \cite{Buchalla:1995vs}
 \begin{align}\label{hsmb}
 \mathcal H_{\rm eff}={\frac{G_F}{\sqrt 2} }
 \sum_{q=d,s}V_{cb}V_{uq}^*\left[C_1(\mu)O_1(\mu)+C_2(\mu)O_2(\mu)\right]+h.c.,
 \end{align}
where the tree operators are
\begin{eqnarray}
O_1=(\bar{q}_{\alpha}u_{\beta})_{V-A}
(\bar{c}_{\beta}b_{\alpha})_{V-A},\qquad
O_2=(\bar{q}_{\alpha}u_{\alpha})_{V-A}
(\bar{c}_{\beta}b_{\beta})_{V-A}.
\end{eqnarray}
In the $SU(3)$ picture, the effective Hamiltonian of charm decay can be written as
\begin{equation}\label{h}
 \mathcal H_{\rm eff}=\sum_{i,j=1}^3 H^{i}_{j}O^{i}_{j},
\end{equation}
where $O^{i}_{j}$ denotes the four-quark operator, and $H$ is the $3\times 3$ coefficient matrix.
The initial and final states of weak decay, such as light meson octet, can be written as
\begin{align}
  |M_\alpha\rangle = (M_\alpha)^{i}_{j}|M^{i}_{j} \rangle,
\end{align}
where $|M^{i}_{j} \rangle$ is the quark composition of meson state, and $(M_\alpha)$ is the coefficient matrix.
The decay amplitude of the $\overline B_\gamma\to D_\alpha M_\beta$ mode is constructed as
\begin{align}\label{amp}
\mathcal{A}(\overline B_\gamma\to D_\alpha M_\beta)& = \langle D_\alpha M_\beta |\mathcal{H}_{\rm eff}| \overline B_\gamma\rangle\nonumber\\&~
=\sum_{\omega}\,(D_\alpha)^n\langle D^n|(M_\beta)^l_m\langle M^l_m||H^{j}_kO^{j}_k||(\overline B_\gamma)_i|\overline B_i\rangle\nonumber\\& ~~=\sum_\omega\,\langle D^{n} M^l_m |O^{j}_{k} |\overline B_{i}\rangle \times (D_\alpha)^{n}(M_\beta)_{m}^{l} H^{j}_{k}(\overline B_\gamma)_i\nonumber\\& ~~~= \sum_\omega X_{\omega}(C_\omega)_{\alpha\beta\gamma},
\end{align}
where $X_\omega = \langle D^{n} M^l_m |O^{j}_{k} |\overline B_{i}\rangle$ is the reduced matrix element, and $(C_\omega)_{\alpha\beta\gamma}=(D_\alpha)^{n}(M_\beta)_{m}^{l} H^{j}_{k}(\overline B_\gamma)_i$ is the Clebsch-Gordan (CG) coefficient \cite{Eckart30,Wigner59}.
The isospin lowering operator $I_-$ is
\begin{eqnarray}
 I_-=  \left( \begin{array}{ccc}
   0   & 0  & 0 \\
     1 &   0  & 0 \\
    0 & 0 & 0 \\
  \end{array}\right).
\end{eqnarray}
Operating $I_-^n$ on an object means applying $I_-$ on it $n$ times.
If the effective Hamiltonian \eqref{h} is zero under $I_-^n$, i.e., $I_-^n\,H=0$, it follows that
\begin{equation}\label{rule}
  \langle D_\alpha M_\beta |I_-^n\,\mathcal{H}_{\rm eff}| \overline B_\gamma\rangle = \sum_\omega\,\langle D^{n} M^l_m |O^{j}_{k} |\overline B_{i}\rangle \times (D_\alpha)^{n}(M_\beta)_{m}^{l} (I_-^n\,H)^{j}_{k}(\overline B_\gamma)_i = 0.
\end{equation}
One can apply $I_-^n$ to the initial/final states and expand the results with the initial/final states as bases. The LHS of Eq.~\eqref{rule} is then turned into sum of several decay amplitudes and the RHS of Eq.~\eqref{rule} remains zero.
Then an isospin sum rule is obtained.

\begin{table*}[t!]
\caption{The values of $n$ for which the Hamiltonian operators of $b\to q_1\overline q_2 q_3$ transitions are zero under $I_-^n$. }\label{n}
\begin{tabular}{|c|c|c|c|c|c|c|c|c|}\hline
 ~~~Mode~~~ &   ~~~$b\to c\overline u d$~~~    & ~~~$b\to c\overline u s$~~~ &  ~~~$b\to c\overline cd$~~~ & ~~~$b\to c\overline cs$~~~ &  ~~~$b\to u\overline u d$~~~    & ~~~$b\to u\overline u s$~~~ &  ~~~$b\to u\overline cd$~~~ & ~~~$b\to u\overline cs$~~~ \\\hline
$n$ & $\geq3$ & $\geq2$ & $\geq2$ & $\geq1$ & $\geq3$ & $\geq2$ & $\geq2$ & $\geq1$ \\\hline
\end{tabular}
\end{table*}
It is found in Ref.~\cite{Wang:2023pnb} that the effective Hamiltonian operators for the $b\to c\overline u d$, $b\to c\overline u s$, $b\to u\overline u d$ and $b\to u\overline u s$ transitions are zero under $I_-^n$ with $n\geq3$, $n\geq2$, $n\geq3$ and $n\geq2$, respectively.
The same trick can be applied to the $b\to c\overline c d$, $b\to c\overline c s$, $b\to u\overline c d$ and $b\to u\overline c s$ transitions.
We find the effective Hamiltonian operators of the four transitions are zero under $I_-^n$ with $n\geq2$, $n\geq1$, $n\geq2$ and $n\geq1$, respectively.
The values of $n$ for which $I_-^n\,H=0$ for all types of transitions are listed in Table.~\ref{n}.

Now let us derive the sum of decay amplitudes generated by $I_-$ for the two- and three-body $\overline B$ decays.
The $\overline B$ meson anti-triplet is defined as
 $|\overline B\rangle = ( |B^-\rangle,\,\, |\overline B^0\rangle ,\,\, |\overline B^0_s\rangle )$.
The charm meson anti-triplet and triplet are $|D\rangle = ( |D^0\rangle,\,\, |D^+\rangle ,\,\, |D^+_s\rangle )$ and $|\overline D\rangle = ( |\overline D^0\rangle,\,\, |D^-\rangle ,\,\, |D^-_s\rangle )$, respectively.
The pseudoscalar meson octet $|M_8 \rangle$ is expressed as
\begin{eqnarray}\label{a1}
 |M_8\rangle =  \left( \begin{array}{ccc}
   \frac{1}{\sqrt 2} |\pi^0\rangle +  \frac{1}{\sqrt 6} |\eta_8\rangle,    & |\pi^+\rangle,  & |K^+\rangle \\
   | \pi^-\rangle, &   - \frac{1}{\sqrt 2} |\pi^0\rangle+ \frac{1}{\sqrt 6} |\eta_8\rangle,   & |K^0\rangle \\
   | K^- \rangle,& |\overline K^0\rangle, & -\sqrt{2/3}|\eta_8\rangle \\
  \end{array}\right).
\end{eqnarray}
And $J/\Psi$ is the $c\overline c$ singlet.
The coefficient matrix of $I_-
\,|\overline B_\alpha\rangle$ expanded by three $\overline B$ meson states, $[I_-]_{\overline B}$, is derived to be \cite{Wang:2023pnb}
\begin{eqnarray}\label{MB}
 [I_-]_{\overline B}= \left( \begin{array}{ccc}
   0   & 0  & 0 \\
     1 &  0  & 0 \\
    0 & 0 & 0 \\
  \end{array}\right).
\end{eqnarray}
The coefficient matrix $[I_-]_{ D}$ is derived to be \cite{Wang:2023pnb}
\begin{eqnarray}\label{MD}
 [I_-]_{ D}= \left( \begin{array}{ccc}
   0   & 1  & 0 \\
     0 &  0  & 0 \\
    0 & 0 & 0 \\
  \end{array}\right),
\end{eqnarray}
and $[I_-]_{ \overline D} = [I_-]_{D}^T$, where superscript "$T$" means transposition of the matrix.
The difference between Eq.~\eqref{MB} and Eq.~\eqref{MD} arises from the fact that the anti-triplet in initial state can be seen as triplet in final state.
If the bases of pseudoscalar meson octet are defined as
\begin{align}
 \langle [M_8]_\alpha| = ( \langle \pi^+|,\,\,\langle \pi^0|,\,\,\langle \pi^-|,\,\,\langle K^+|,\,\,\langle K^0|,\,\,\langle \overline K^0|,\,\,\langle K^-|,\,\,\langle \eta_8|    ),
\end{align}
the coefficient matrix $[I_-]_{M_8}$ is \cite{Wang:2023pnb}.
\begin{eqnarray}\label{MP}
 [I_-]_{M_8}= \left( \begin{array}{cccccccc}
  0 & 0& 0& 0& 0& 0& 0& 0 \\
  -\sqrt{2}& 0& 0& 0& 0& 0& 0& 0 \\
 0& \sqrt{2}& 0& 0& 0& 0& 0& 0 \\
  0& 0& 0& 0& 0& 0& 0& 0 \\
  0& 0& 0& 1& 0& 0& 0& 0\\
 0& 0& 0& 0& 0& 0& 0& 0\\
 0& 0& 0& 0& 0& -1& 0& 0 \\
 0& 0& 0& 0& 0&0& 0& 0 \\
  \end{array}\right).
\end{eqnarray}
And $I_-\,|J/\Psi\rangle$ is zero.
With coefficient matrices \eqref{MB}, \eqref{MD} and \eqref{MP}, the sum of decay amplitudes generated by $I_-$ for the $\overline B \to DM$ mode is written as
\begin{align}\label{rulex1}
{ SumI_-}\,[\gamma, \alpha,\beta]= \sum_\mu\left[-\{[I_-]_{D}\}_\alpha^\mu \mathcal{A}_{ \gamma \to \mu \beta} +  \{[I_-]_{M_8}\}_\beta^\mu \mathcal{A}_{\gamma\to \alpha\mu } + \{[I_-]_{\overline B}\}_\gamma^\mu \mathcal{A}_{\mu\to \alpha \beta }\right],
\end{align}
in which
$\mathcal{A}_{ \gamma \to \mu \beta}$, $\mathcal{A}_{\gamma\to \alpha\mu }$, $\mathcal{A}_{\mu\to \alpha \beta }$ are the decay amplitudes of $\overline B_\gamma\to D_\mu M_\beta$, $\overline B_\gamma\to D_\alpha M_\mu$, $\overline B_\mu\to D_\alpha M_\beta$ respectively.
The minus sign in the first term matches the minus sign in $I_-$ acting on meson octet, $I_-\langle M_8 | = I_- \cdot \langle M_8 | - \langle M_8 |\cdot I_-$.
One can apply Eq.~\eqref{rulex1} three (two) or more times with appropriate $\alpha$, $\beta$, $\gamma$ to get an isospin sum rule for $b\to c\overline u d$ ($b\to c\overline u s$) modes.

Similarly, the sum of decay amplitudes for the $\overline B\to D\overline D$ modes generated by $I_-$ is
\begin{align}\label{rulex2}
 { SumI_-}\,[\,\gamma, \alpha,\beta\,]= &  \sum_\mu\Big[-\{[I_-]_{D}\}_\alpha^\mu \mathcal{A}_{ \gamma \to \mu \beta} +  \{[I_-]_{\overline D}\}_\beta^\mu \mathcal{A}_{\gamma\to \alpha\mu }+ \{[I_-]_{\overline B}\}_\gamma^\mu \mathcal{A}_{\mu\to \alpha \beta }\Big].
\end{align}
The sum of decay amplitudes of $\overline B\to J/\Psi M$ mode generated by $I_-$ is
\begin{align}\label{rulex3}
 { SumI_-}\,[\,\gamma, \alpha,\beta\,]=  \sum_\mu\Big[ \{[I_-]_{M_8}\}_\beta^\mu \mathcal{A}_{\gamma\to \alpha\mu } + \{[I_-]_{\overline B}\}_\gamma^\mu \mathcal{A}_{\mu\to \alpha \beta}\Big].
\end{align}
The sum of decay amplitudes generated by $I_-$ for the $\overline B\to M M$ mode is
\begin{align}\label{rulex4}
 { SumI_-}\,[\gamma, \alpha,\beta]=  \sum_\mu\left[\{[I_-]_{M_8}\}_\alpha^\mu \mathcal{A}_{ \gamma \to \mu \beta} +  \{[I_-]_{M_8}\}_\beta^\mu \mathcal{A}_{\gamma\to \alpha\mu } + \{[I_-]_{\overline B}\}_\gamma^\mu \mathcal{A}_{\mu\to \alpha \beta }\right].
\end{align}
And the sum of decay amplitudes generated by $I_-$ for the $\overline B\to \overline D M$ mode is
\begin{align}\label{rulex5}
 { SumI_-}\,[\gamma, \alpha,\beta]=  \sum_\mu\left[\{[I_-]_{\overline D}\}_\alpha^\mu \mathcal{A}_{ \gamma \to \mu \beta} +  \{[I_-]_{M_8}\}_\beta^\mu \mathcal{A}_{\gamma\to \alpha\mu } + \{[I_-]_{\overline B}\}_\gamma^\mu \mathcal{A}_{\mu\to \alpha \beta }\right].
\end{align}

Above master formulas can be extended to the three-body decays.
The sum of decay amplitudes of $\overline B\to DMM$ mode generated by $I_-$ is
\begin{align}\label{rule1}
 { SumI_-}\,[\,\delta, \alpha,\beta,\gamma\,]= &  \sum_\mu\Big[-\{[I_-]_{D}\}_\alpha^\mu \mathcal{A}_{ \delta \to \mu \beta\gamma} +  \{[I_-]_{M_8}\}_\beta^\mu \mathcal{A}_{\delta\to \alpha\mu\gamma }\nonumber\\&~~~~~ + \{[I_-]_{M_8}\}_\gamma^\mu \mathcal{A}_{\delta\to \alpha\beta\mu }+ \{[I_-]_{\overline B}\}_\delta^\mu \mathcal{A}_{\mu\to \alpha \beta \gamma}\Big].
\end{align}
The sum of decay amplitudes of $\overline B\to D\overline DM$ mode generated by $I_-$ is
\begin{align}\label{rule2}
 { SumI_-}\,[\,\delta, \alpha,\beta,\gamma\,]= &  \sum_\mu\Big[-\{[I_-]_{D}\}_\alpha^\mu \mathcal{A}_{ \delta \to \mu \beta\gamma} +  \{[I_-]_{\overline D}\}_\beta^\mu \mathcal{A}_{\delta\to \alpha\mu\gamma } \nonumber\\&~~~~~ + \{[I_-]_{M_8}\}_\gamma^\mu \mathcal{A}_{\delta\to \alpha\beta\mu }+ \{[I_-]_{\overline B}\}_\delta^\mu \mathcal{A}_{\mu\to \alpha \beta \gamma}\Big].
\end{align}
The sum of decay amplitudes of $\overline B\to J/\Psi MM$ mode generated by $I_-$ is
\begin{align}\label{rule3}
 { SumI_-}\,[\,\delta, \alpha,\beta,\gamma\,]=  \sum_\mu\Big[ \{[I_-]_{M_8}\}_\beta^\mu \mathcal{A}_{\delta\to \alpha\mu\gamma } + \{[I_-]_{M_8}\}_\gamma^\mu \mathcal{A}_{\delta\to \alpha\beta\mu }+ \{[I_-]_{\overline B}\}_\delta^\mu \mathcal{A}_{\mu\to \alpha \beta \gamma}\Big].
\end{align}
The sum of decay amplitudes of $\overline B\to MMM$ mode generated by $I_-$ is
\begin{align}\label{rule4}
 { SumI_-}\,[\,\delta, \alpha,\beta,\gamma\,]= &  \sum_\mu\Big[\{[I_-]_{M_8}\}_\alpha^\mu \mathcal{A}_{ \delta \to \mu \beta\gamma} +  \{[I_-]_{M_8}\}_\beta^\mu \mathcal{A}_{\delta\to \alpha\mu\gamma }\nonumber\\&~~~~~ + \{[I_-]_{M_8}\}_\gamma^\mu \mathcal{A}_{\delta\to \alpha\beta\mu }+ \{[I_-]_{\overline B}\}_\delta^\mu \mathcal{A}_{\mu\to \alpha \beta \gamma}\Big].
\end{align}
The sum of decay amplitudes of $\overline B\to \overline DMM$ mode generated by $I_-$ is
\begin{align}\label{rule5}
 { SumI_-}\,[\,\delta, \alpha,\beta,\gamma\,]= &  \sum_\mu\Big[\{[I_-]_{\overline D}\}_\alpha^\mu \mathcal{A}_{ \delta \to \mu \beta\gamma} +  \{[I_-]_{M_8}\}_\beta^\mu \mathcal{A}_{\delta\to \alpha\mu\gamma } \nonumber\\&~~~~~ + \{[I_-]_{M_8}\}_\gamma^\mu \mathcal{A}_{\delta\to \alpha\beta\mu }+ \{[I_-]_{\overline B}\}_\delta^\mu \mathcal{A}_{\mu\to \alpha \beta \gamma}\Big].
\end{align}

The isospin sum rules derived from Eqs.~\eqref{rulex1}$\sim$\eqref{rule5} are listed in Appendix \ref{rules}.
Many new isospin sum rules, which are not present in literature such as \cite{Savage:1989ub}, are derived.
Especially those isospin sum rules involving three or more decay channels are derived for the first time.
One can check these isospin sum rules by writing the isospin amplitudes.
It is important to note that the order of final states in the isospin sum rules for three-body decays cannot be exchanged arbitrarily, or else the isospin relations of intermediate resonance strong decays are broken \cite{Wang:2023don}.

The decay modes dominated by the $b\to c\overline cd/s$ ($b\to u\overline ud/s$) transitions also receive contributions from the $b\to u\overline ud/s$ ($b\to c\overline cd/s$) transitions via annihilation topologies.
In the derivation of the isospin sum rules, the leading CKM matrix contributions should be considered.
The isospin sum rules derived from $I^n_- \,H_{u\overline ud/s}=0$ are not broken by the $b\to c\overline cd/s$ transitions because if $I^n_- \,H_{u\overline ud/s}=0$, $I^n_-\, H_{ c\overline cd/s}$ must be zero, as  indicated in Table.~\ref{n}.
However, the isospin sum rules derived from $I^n_- \,H_{c\overline cd/s}=0$ are broken by the $b\to u\overline ud/s$ contributions.
The breaking induced by $I^n_-\, H_{ u\overline us}\neq 0$ is naively predicted to be $ |\frac{V_{ub}V_{us}}{V_{cb}V_{cs}}|\sim
\mathcal{O}(1\%)$, while the breaking induced by $I^n_-\, H_{ u\overline ud}\neq 0$ is naively predicted to be $ |\frac{V_{ub}V_{ud}}{V_{cb}V_{cd}}|\sim
\mathcal{O}(10\%)$.
Thus, we discard the isospin sum rules derived from $I^n_- \,H_{c\overline cd}=0$ but $I^n_- \,H_{u\overline ud}\neq0$.
Among the isospin sun rules listed in \ref{cc},
only Eqs.~\eqref{cc1}, \eqref{r1}, \eqref{r3}, \eqref{cc2}, \eqref{cc3}, and \eqref{r5} hold without the approximation $V_{ub}V_{us}\ll V_{cb}V_{cs}$.
Most of them are broken at around $\mathcal{O}(1\%)$ due to sub-leading CKM corrections.

\section{Phenomenological analysis}\label{PA}
\begin{table*}[t!]
\caption{Testing the isospin sum rules involving two decay channels, in which $f$ is the scaling factor between the two decay amplitudes in one isospin sum rule. All the data are taken from PDG \cite{PDG}. Branching fractions of the $B^0\to J/\Psi K^0\eta$, $B^0\to J/\Psi K^0\omega$ and $B^0\to J/\Psi K^0\phi$ are taken from branching fractions of the $B^0\to J/\Psi K^0_S\eta$, $B^0\to J/\Psi K^0_S\omega$ and $B^0\to J/\Psi K^0_S\phi$ by neglecting the DCS transitions.}\label{testd}
{\footnotesize
\begin{tabular}{|c|c|c|c|c|c|}\hline\hline
~~~~Channel~~~~ &~~$\mathcal{B}r_{\rm exp}$~~ & ~~~~Channel~~~~ & ~~$\mathcal{B}r_{\rm exp}$~~ & ~Sum rule~ & ~~$f\times|\mathcal{A}_1/\mathcal{A}_2|$~~\\
\hline
$B^+\to  \overline D^0D^+_s$ &$(9.0\pm 0.9)\times 10^{-3}$& $B^0\to D^-D^+_s$& $(7.2\pm 0.8)\times 10^{-3}$ & \eqref{test3}& $1.02\pm 0.08$\\ \hline
$B^+\to  \overline D^0D^{*+}_s$ &$(7.6\pm 1.6)\times 10^{-3}$& $B^0\to D^-D^{*+}_s$& $(7.4\pm 1.6)\times 10^{-3}$ & \eqref{test3}& $0.92\pm0.14$\\ \hline
$B^+\to  \overline D^{*0}D^+_s$ &$(8.2\pm 1.7)\times 10^{-3}$& $B^0\to D^{*-}D^+_s$& $(8.0\pm 1.1)\times 10^{-3}$ & \eqref{test3}& $0.92\pm0.12$\\ \hline
$B^+\to  \overline D^{*0}D^{*+}_s$ &$(1.71\pm 0.24)\%$& $B^0\to D^{*-}D^{*+}_s$& $(1.77\pm 0.14)\%$ & \eqref{test3}& $0.89\pm0.08$\\ \hline
$B^0_s\to  D^+D^-$ &$(2.2\pm 0.6)\times 10^{-4}$& $B^0_s\to D^0\overline D^0$& $(1.9\pm 0.5)\times 10^{-4}$ & \eqref{test7}& $1.08\pm 0.21$\\ \hline
$B^+\to  J/\Psi K^+$ &$(1.020\pm 0.019)\times 10^{-3}$& $B^0\to J/\Psi K^0$& $(0.891\pm 0.021)\times 10^{-3}$ & \eqref{test5}& $0.97\pm0.02$\\ \hline
$B^+\to  J/\Psi K^{*+}$ &$(1.43\pm 0.08)\times 10^{-3}$& $B^0\to J/\Psi K^{*0}$& $(1.27\pm 0.05)\times 10^{-3}$ & \eqref{test5}& $0.96\pm 0.04$\\ \hline
$B^+\to  D_s^+ \pi^0$ &$(1.6\pm 0.5)\times 10^{-5}$& $B^0\to D_s^+ \pi^-$& $(2.03\pm 0.18)\times 10^{-5}$ & \eqref{test19}& $1.14\pm 0.19$\\ \hline
$B^+\to  J/\Psi K^+\eta$ &$(1.24\pm 0.14)\times 10^{-4}$& $B^0\to J/\Psi K^0\eta$& $(1.08\pm 0.18)\times 10^{-4}$ & \eqref{test6}& $0.97\pm 0.10$\\ \hline
$B^+\to  J/\Psi K^+\omega$ &$(3.20^{+0.60}_{-0.32})\times 10^{-4}$& $B^0\to J/\Psi K^0\omega$& $(2.3\pm 0.4)\times 10^{-4}$ & \eqref{test6}& $1.07^{+0.14}_{-0.11}$\\ \hline
$B^+\to  J/\Psi K^+\phi$ &$(5.0\pm 0.4)\times 10^{-5}$& $B^0\to J/\Psi K^0\phi$& $(4.9\pm 1.0)\times 10^{-5}$ & \eqref{test6}& $0.92\pm 0.10$\\ \hline
$B^+\to  D^-D^+ K^+$ &$(2.2\pm 0.7)\times 10^{-4}$& $B^0\to \overline D^0D^0K^0$& $(2.7\pm 1.1)\times 10^{-4}$ & \eqref{test11}& $0.82\pm 0.22 $\\ \hline
$B^+\to  D^{*-}D^{*+} K^+$ &$(1.32\pm 0.18)\times 10^{-3}$& $B^0\to \overline D^{*0}D^{*0}K^0$& $(2.4\pm 0.9)\times 10^{-3}$ & \eqref{test11}& $0.67\pm 0.14$\\ \hline
$B^+\to  \overline D^0D^+ K^0$ &$(1.55\pm 0.21)\times 10^{-3}$& $B^0\to  D^-D^0K^+$& $(1.07\pm 0.11)\times 10^{-3}$ & \eqref{test12}& $1.09\pm 0.10$\\ \hline
$B^+\to  \overline D^0D^{*+} K^0$ &$(3.8\pm 0.4)\times 10^{-3}$& $B^0\to  D^-D^{*0}K^+$& $(3.5\pm 0.4)\times 10^{-3}$ & \eqref{test12}& $0.95\pm 0.08$\\ \hline
$B^+\to  \overline D^{*0}D^{+} K^0$ &$(2.1\pm 0.5)\times 10^{-3}$& $B^0\to  D^{*-}D^{0}K^+$& $(2.47\pm 0.21)\times 10^{-3}$ & \eqref{test12}& $0.84\pm 0.11$\\ \hline
$B^+\to  \overline D^{*0}D^{*+} K^0$ &$(9.2\pm 1.2)\times 10^{-3}$& $B^0\to  D^{*-}D^{*0}K^+$& $(10.6\pm 0.9)\times 10^{-3}$ & \eqref{test12}& $0.85\pm 0.07$\\ \hline
$B^+\to  \overline D^{0}D^{0} K^+$ &$(1.45\pm 0.33)\times 10^{-3}$& $B^0\to  D^{-}D^{+}K^0$& $(7.5\pm 1.7)\times 10^{-4}$ & \eqref{test13}& $1.26\pm 0.21$\\ \hline
$B^+\to  \overline D^{*0}D^{*0} K^+$ &$(11.2\pm 1.3)\times 10^{-3}$& $B^0\to  D^{*-}D^{*+}K^0$& $(8.1\pm 0.7)\times 10^{-3}$ & \eqref{test13}& $1.07\pm 0.08$\\ \hline
$B^+\to  J/\Psi\pi^+ K^0$ &$(1.14\pm 0.11)\times 10^{-3}$& $B^0\to  J/\Psi\pi^- K^+$& $(1.15\pm 0.05)\times 10^{-3}$ & \eqref{test16}& $0.90\pm 0.05$\\ \hline
\end{tabular}}
\end{table*}

In this section, we discuss physical applications of the isospin sum rules.
The isospin sum rules analyzed in this section are listed below.
\begin{align}\label{test3}
\mathcal{A}(B^-\to D^0D^-_s)=\mathcal{A}( \overline B^0\to D^+
 D^-_s),
\end{align}
\begin{align}\label{test7}
\mathcal{A}(\overline B^0_s\to D^0\overline D^0)=\mathcal{A}( \overline B^0_s\to D^+
 D^-),
\end{align}
\begin{align}\label{test5}
\mathcal{A}( B^-\to J/\Psi K^-)=\mathcal{A}( \overline B^0\to J/\Psi\overline K^0),
\end{align}
\begin{align}\label{test18}
\mathcal{A}(\overline B^0_s\to \pi^0\pi^0)=\mathcal{A}(\overline B^0_s\to \pi^+\pi^-),
\end{align}
\begin{align}\label{test19}
\mathcal{A}( \overline B^0\to D^-_s\pi^+)=\sqrt{2}\,\mathcal{A}( B^-\to D^-_s\pi^0),
\end{align}
\begin{align}\label{test6}
\mathcal{A}( B^-\to J/\Psi K^-\eta_8)=\mathcal{A}(\overline B^0\to J/\Psi \overline K^0\eta_8),
\end{align}
\begin{align}\label{test17}
\mathcal{A}( \overline B^0_s\to J/\Psi K^0 \overline K^0)=\mathcal{A}( \overline B^0_s\to J/\Psi K^+ K^-),
\end{align}
\begin{align}\label{test8}
\mathcal{A}(B^-\to D^0\pi^-)+\sqrt{2}\,\mathcal{A}(\overline B^0\to D^0\pi^0)-\mathcal{A}(\overline B^0\to D^+\pi^-)=0,
\end{align}
\begin{align}\label{test1}
\mathcal{A}(B^-\to D^0K^-)-\,\mathcal{A}(\overline B^0\to D^0\overline K^0)-\mathcal{A}(\overline B^0\to D^+K^-)=0,
\end{align}
\begin{align}\label{test9}
\sqrt{2}\,\mathcal{A}(B^-\to \pi^0\pi^-)+\,\mathcal{A}(\overline B^0\to \pi^0\pi^0)-\mathcal{A}(\overline B^0\to \pi^+\pi^-)=0,
\end{align}
\begin{align}\label{test10}
\mathcal{A}( B^-\to D^-\overline K^0)-\mathcal{A}( B^-\to \overline D^0 K^-)+\mathcal{A}( \overline B^0\to  \overline D^0\overline K^0)=0.
\end{align}
All these isospin sum rules can be found in Appendix \ref{rules}.
In fact, not all the isospin sum rules listed in Appendix \ref{rules} are independent.
The five isospin sum rules \eqref{r1}$\sim$\eqref{r2} imply four independent isospin sum rules,
\begin{align}\label{test11}
\mathcal{A}(B^-\to D^+D^-K^-)=\mathcal{A}(\overline B^0\to D^0
 \overline D^0\overline K^0),
\end{align}
\begin{align}\label{test12}
\mathcal{A}(B^-\to D^0D^-\overline K^0)=\mathcal{A}(\overline B^0\to D^+
 \overline D^0 K^-),
\end{align}
\begin{align}\label{test13}
\mathcal{A}(B^-\to D^0\overline D^0K^-)=\mathcal{A}(\overline B^0\to D^+
  D^- \overline K^0),
\end{align}
\begin{align}\label{test14}
\mathcal{A}(B^-\to D^0D^-\overline K^0)-\mathcal{A}(B^-\to D^0
\overline D^0K^-)+\mathcal{A}(B^-\to D^+
 D^-K^-)=0.
\end{align}
The four isospin sum rules \eqref{r5}$\sim$\eqref{r6} imply three independent isospin sum rules,
\begin{align}\label{test16}
&\mathcal{A}( B^-\to J/\Psi\pi^- \overline K^0)=\sqrt{2}\,\mathcal{A}( B^-\to J/\Psi\pi^0 K^-)\nonumber\\&~~~~~~=\mathcal{A}(\overline B^0\to J/\Psi\pi^+ K^-)=-\sqrt{2}\,\mathcal{A}(\overline B^0\to J/\Psi\pi^0 \overline K^0).
\end{align}
And the four isospin sum rules \eqref{r3}$\sim$\eqref{r4} imply three independent isospin sum rules,
\begin{align}\label{test15}
&\mathcal{A}(B^-\to D^+D^-_s \pi^-)=\sqrt{2}\,\mathcal{A}(B^-\to D^0D^-_s \pi^0)\nonumber\\&~~~~~~=\mathcal{A}(\overline B^0\to D^0D^-_s \pi^+)=-\sqrt{2}\,\mathcal{A}(\overline B^0\to D^+D^-_s \pi^0)=0.
\end{align}
The final states of decay channels in above isospin sum rules can be replaced by their excited states.

If two decay channels form an isospin sum rule, the branching fractions of the two decay channels have a definite proportional relation.
With the data given by Particle Data Group \cite{PDG}, we test the isospin symmetry, as shown in Table.~\ref{testd}.
If isospin symmetry holds precise, the ratio of the decay amplitude multiplied by the scaling factor, $f\times|\mathcal{A}_1/\mathcal{A}_2|$ in Table.~\ref{testd}, is unit.
One can find the isospin breaking in most decay modes is consistent with the naive prediction $\delta_I\sim \mathcal{O}(1\%)$, except for some modes involving two vector mesons.
The isospin breaking in the $B^+\to  D^{*-}D^{*+} K^+$ and $B^0\to \overline D^{*0}D^{*0}K^0$ modes is $\mathcal{O}(10\%)$ with a significance of $2.4\,\sigma$.
Even the sub-leading CKM contributions are considered, the isospin breaking in this case is much larger than the naive prediction.
More precise measurements could verify the abnormal isospin breaking.
On the other hand, some branching fractions of $B$ meson decays can be estimated using the isospin sum rules involving two decay channels.
Our results are presented in Table \ref{pred}.
Future experiments can test the isospin symmetry by measuring these predicted branching fractions.

\begin{table*}[t!]
\caption{Predictions for branching fractions of the $B$ meson two- or three-body weak decays, in which all data are taken from PDG \cite{PDG}. }\label{pred}
\begin{tabular}{|c|c|c|c|c|}\hline\hline
~~~~~Channel~~~~~ & ~~~~$\mathcal{B}r_{\rm exp}$~~~~ & ~~~~~Channel~~~~~ & ~~~~$\mathcal{B}r_{\rm th}$~~~~ & ~~Sum rule~~ \\
\hline
$B^0_s\to  \pi^+\pi^-$ &$(7.0\pm 1.0)\times 10^{-7}$& $B^0_s\to  \pi^0\pi^0$& $(7.0\pm 1.0)\times 10^{-7}$ & \eqref{test18}\\ \hline
$B^0\to  D^{*+}_s\pi^-$ &$(2.1\pm 0.4)\times 10^{-5}$& $B^+\to  D^{*+}_s\pi^0$& $(1.3\pm 0.2)\times 10^{-5}$ & \eqref{test19}\\ \hline
$B^0\to  D^{*+}_s\rho^-$ &$(4.1\pm 1.3)\times 10^{-5}$& $B^+\to  D^{*+}_s\rho^0$& $(2.5\pm 0.8)\times 10^{-5}$ & \eqref{test19}\\ \hline
$B^0_s\to  J/\Psi K^+K^-$ &$(7.9\pm 0.7)\times 10^{-4}$& $B^0_s\to  J/\Psi K^0\overline K^0$& $(7.9\pm 0.7)\times 10^{-4}$ & \eqref{test17}\\ \hline
~$B^0_s\to  J/\Psi K^{*0}\overline K^{*0}$~ &~$(1.10\pm 0.09)\times 10^{-4}$~& ~$B^0_s\to  J/\Psi K^{*+}K^{*-}$~& ~$(1.10\pm 0.09)\times 10^{-4}$~ & \eqref{test17}\\ \hline
$B^+\to  D^{*-}D^{+}K^+$ &$(6.0\pm 1.3)\times 10^{-4}$& $B^0\to \overline D^{*0}D^{0}K^+$& $(5.0\pm 1.1)\times 10^{-4}$ & \eqref{test11}\\ \hline
~$B^+\to  D^-D^{*+}K^+$~ &~$(6.3\pm 1.1)\times 10^{-4}$~& ~$B^0\to \overline D^{0}D^{*0}K^+$~& ~$(5.2\pm 0.9)\times 10^{-4}$~ & \eqref{test11}\\ \hline
$B^+\to  \overline D^{*0}D^{0}K^+$ &$(2.26\pm 0.23)\times 10^{-3}$& $B^0\to D^{*-}D^{+}K^0$& $(1.87\pm 0.19)\times 10^{-3}$ & \eqref{test13}\\ \hline
$B^+\to  \overline D^{0}D^{*0}K^+$ &$(6.3\pm 0.5)\times 10^{-3}$& $B^0\to D^{-}D^{*+}K^0$& $(5.2\pm 0.4)\times 10^{-3}$ & \eqref{test13}\\ \hline
$B^0\to  J/\Psi\pi^- K^+$ &$(1.15\pm 0.05)\times 10^{-3}$& $B^+\to  J/\Psi\pi^0 K^+$& $(7.0\pm 0.3)\times 10^{-4}$ & \eqref{test16}\\ \hline
$B^0\to  J/\Psi\pi^- K^+$ &$(1.15\pm 0.05)\times 10^{-3}$& $B^0\to  J/\Psi\pi^0 K^0$& $(5.8\pm 0.3)\times 10^{-4}$ & \eqref{test16}\\ \hline
\end{tabular}
\end{table*}
\begin{table*}[t!]
\caption{Testing the isospin sum rules involving two decay channels, where "Yes" denote the isospin triangle is closed and "No" denote the isospin triangle fails to be closed. All the data are taken from PDG \cite{PDG}. }\label{testt}
{\scriptsize
\begin{tabular}{|c|c|c|c|c|c|}\hline\hline
~~~Channel~~~ & $\mathcal{B}r_{\rm exp}$ & ~~~Channel~~~ & $\mathcal{B}r_{\rm exp}$ & ~~~Channel~~~ & $\mathcal{B}r_{\rm exp}$\\
\hline
$B^+\to \overline D^0\pi^+$ &$(4.61\pm 0.10)\times 10^{-3}$& $B^0\to D^-\pi^+$& $(2.51\pm 0.08)\times 10^{-3}$ &  $B^0\to \overline D^0\pi^0$& $(2.67\pm 0.09)\times 10^{-4}$ \\ \hline
 \multicolumn{3}{|c|}{Sum rule\quad\eqref{test8}} & \multicolumn{3}{c|}{Isospin triangle\quad $\Rightarrow$\quad Yes}  \\\hline
$B^+\to \overline D^0\rho^+$ &$(1.34\pm 0.18)\%$& $B^0\to D^-\rho^+$& $(7.6\pm 1.2)\times 10^{-3}$ &  $B^0\to \overline D^0\rho^0$& $(3.21\pm 0.21)\times 10^{-4}$ \\ \hline
 \multicolumn{3}{|c|}{Sum rule\quad\eqref{test8}} & \multicolumn{3}{c|}{Isospin triangle\quad $\Rightarrow$\quad Yes}  \\\hline
$B^+\to \overline D^0 K^+$ &$(3.64\pm 0.15)\times 10^{-4}$& $B^0\to D^- K^+$& $(2.05\pm 0.08)\times 10^{-4}$ &  $B^0\to \overline D^0 K^0$& $(5.2\pm 0.7)\times 10^{-5}$ \\ \hline
 \multicolumn{3}{|c|}{Sum rule\quad\eqref{test1}} & \multicolumn{3}{c|}{Isospin triangle\quad $\Rightarrow$\quad Yes}  \\\hline
$B^+\to \overline D^0 K^{*+}$ &$(5.3\pm 0.4)\times 10^{-4}$& $B^0\to D^- K^{*+}$& $(4.5\pm 0.7)\times 10^{-4}$ &  $B^0\to \overline D^0 K^{*0}$& $(4.5\pm 0.6)\times 10^{-5}$ \\ \hline
 \multicolumn{3}{|c|}{Sum rule\quad\eqref{test1}} & \multicolumn{3}{c|}{Isospin triangle\quad $\Rightarrow$\quad Yes}  \\\hline
$B^+\to \pi^+ \pi^0$ &$(5.5\pm 0.4)\times 10^{-6}$& $B^0\to \pi^+ \pi^-$& $(5.12\pm 0.19)\times 10^{-6}$ &  $B^0\to \pi^0 \pi^{0}$& $(1.59\pm 0.26)\times 10^{-6}$ \\ \hline
 \multicolumn{3}{|c|}{Sum rule\quad\eqref{test9}} & \multicolumn{3}{c|}{Isospin triangle\quad $\Rightarrow$\quad Yes}  \\\hline
$B^+\to \rho^+ \rho^0$ &$(2.40\pm 0.19)\times 10^{-5}$& $B^0\to \rho^+ \rho^-$& $(2.77\pm 0.19)\times 10^{-5}$ &  $B^0\to \rho^0 \rho^{0}$& $(0.96\pm 0.15)\times 10^{-6}$ \\ \hline
 \multicolumn{3}{|c|}{Sum rule\quad\eqref{test9}} & \multicolumn{3}{c|}{Isospin triangle\quad $\Rightarrow$\quad No}  \\\hline
$B^+\to D^0 K^{*+}$ &$(5.4^{+1.8}_{-4.0})\times 10^{-6}$& $B^+\to D^+ K^{*0}$& $<4.9\times 10^{-7}$ &  $B^0\to D^0 K^{*0}$& $(3.0\pm 0.6)\times 10^{-6}$ \\ \hline
 \multicolumn{3}{|c|}{Sum rule\quad\eqref{test10}} & \multicolumn{3}{c|}{Isospin triangle\quad $\Rightarrow$\quad Yes}  \\\hline
$B^+\to \overline D^0 D^+ K^{0}$ &$(1.55\pm 0.21)\times 10^{-3}$& $B^+\to \overline D^0 D^0 K^{+}$& $(1.45\pm 0.33)\times 10^{-3}$ &  $B^+\to D^-D^+ K^{+}$& $(2.2\pm 0.7)\times 10^{-4}$ \\ \hline
 \multicolumn{3}{|c|}{Sum rule\quad\eqref{test14}} & \multicolumn{3}{c|}{Isospin triangle\quad $\Rightarrow$\quad Yes}  \\\hline
$B^+\to \overline D^{*0} D^+ K^{0}$ &$(2.1\pm 0.5)\times 10^{-3}$& $B^+\to \overline D^{*0} D^0 K^{+}$& $(2.26\pm 0.23)\times 10^{-3}$ &  $B^+\to D^{*-}D^+ K^{+}$& $(6.0\pm 1.3)\times 10^{-4}$ \\ \hline
 \multicolumn{3}{|c|}{Sum rule\quad\eqref{test14}} & \multicolumn{3}{c|}{Isospin triangle\quad $\Rightarrow$\quad Yes}  \\\hline
$B^+\to \overline D^0 D^{*+} K^{0}$ &$(3.8\pm 0.4)\times 10^{-3}$& $B^+\to \overline D^0 D^{*0} K^{+}$& $(6.3\pm 0.5)\times 10^{-3}$ &  $B^+\to D^-D^{*+} K^{+}$& $(6.3\pm 1.1)\times 10^{-4}$ \\ \hline
 \multicolumn{3}{|c|}{Sum rule\quad\eqref{test14}} & \multicolumn{3}{c|}{Isospin triangle\quad $\Rightarrow$\quad Yes}  \\\hline
$B^+\to \overline D^{*0} D^{*+} K^{0}$ &$(9.2\pm 1.2)\times 10^{-3}$& $B^+\to \overline D^{*0} D^{*0} K^{+}$& $(11.2\pm 1.3)\times 10^{-3}$ &  $B^+\to D^{*-}D^{*+} K^{+}$& $(1.32\pm 0.18)\times 10^{-3}$ \\ \hline
 \multicolumn{3}{|c|}{Sum rule\quad\eqref{test14}} & \multicolumn{3}{c|}{Isospin triangle\quad $\Rightarrow$\quad Yes}  \\\hline
\end{tabular}}
\end{table*}
If there are three decay channels forming an isospin sum rule, the decay amplitudes of them form an isospin triangle.
The sum of any two sides of a triangle is greater than the third side, and the difference between any two sides is less than the third side.
We test the isospin sum rules involving three decay channels by checking whether the isospin triangle is closed.
The results are shown in Table.~\ref{testt}.
It is found most of the isospin triangles are closed except for the one formed by the $B^+\to \rho^+\rho^0$, $B^0\to \rho^+\rho^-$ and $B^0\to \rho^0\rho^0$ modes.
It indicates the isospin breaking in the $B\to \rho\rho$ system is larger than the naive prediction.

Form above analysis, it is found the abnormal isospin breaking occurs only in modes involving two vector mesons.
Compared to $B\to PP$ and $PV$ decays, the $B\to VV$ modes have three components of polarization.
Recent QCD studies show the subleading power contributions, which are sensitive to the isospin breaking effects, are very important in exclusive $B\to VV$ decays \cite{Huang:2023jdu,Lu:2022kos}.
Additionally, the final state interactions of $VV$ system have been proved to be larger than the $PP$ and $PV$ systems \cite{Cao:2023csx,Cao:2023gfv}.
It is possible that the isospin breaking in the $B\to VV$ mode is enhanced by the complex dynamics of vector meson system.

$B$ meson decays are a good place for the hidden- and open-charm tetraquark production \cite{An:2022vtg,Chen:2020eyu,Yu:2017pmn}.
Considering the contributions from intermediate resonance, an isospin sum rule for three-body decays involves several isospin sum rules for two-body decays \cite{Wang:2023don}.
If one decay channel is contributed by an intermediate resonance, the rest of channels in the isospin sum rule should be contributed by its isospin partners, or else the isospin symmetry is broken.
Thereby, the isospin sum rules for three-body $B$ meson decays could help us search for the isospin partners of hidden- and open-charm tetraquarks.

The hidden-charm tetraquarks with constituents $c\overline c u\overline d$ and $c\overline c s\overline u$ have been observed by experiments in the $\overline B^0 \to J/\Psi \pi^+K^-$ and $B^- \to J/\Psi K^-\phi$ channels \cite{Belle:2014nuw,LHCb:2021uow}.
According to Eqs.~\eqref{test6} and \eqref{test16}, the isospin partners of these two hidden-charm tetraquarks could be found in the $ B^- \to J/\Psi \pi^-\overline K^0$ and $\overline B^0 \to J/\Psi \overline K^0\phi$ channels.
The open-charm tetraquark with constituent $cs \overline u\overline d$ has been observed in the $ B^- \to D^+D^-K^-$ mode \cite{LHCb:2020pxc}.
According to Eq.~\eqref{test11}, the isospin partner of this tetraquark could be found in the $ \overline B^0 \to D^0\overline D^0 \overline K^0$ mode.
The open-charm tetraquarks with constituents $\overline c\overline u d s$ and $\overline c\overline d u s$ were observed in the $B^- \to D^+D^-_s\pi^-$ and $\overline B^0 \to D^0D^-_s\pi^+$ decays \cite{LHCb:2022sfr,LHCb:2022lzp}.
Their isospin partners might be observed in the $\pi^0$ involved modes $B^- \to D^0D^-_s\pi^0$ and $\overline B^0 \to D^+D^-_s\pi^0$ according to Eq.~\eqref{test15}.

\section{Summary}\label{SUM}

Isospin symmetry is the most precise flavor symmetry.
In this work, we derived the master formulas of isospin sum rules for the two- and three-body non-leptonic decays of $B$ mesons.
Many new isospin sum rules were found by these master formulas.
The isospin sum rules can be used to test isospin symmetry and provide hints for the isospin partners of exotic hadrons.
The isospin breaking could reach to be at order of $10\%$ in some decay modes involving two vector mesons.
And the charm tetraquarks might be observed in the $ B^- \to J/\Psi \pi^-\overline K^0$, $\overline B^0 \to J/\Psi \overline K^0\phi$ and $\overline B^0\to D^0\overline D^0\overline K^0$ modes according to the isospin symmetry.

\begin{acknowledgements}

This work was supported in part by the National Natural Science Foundation of China under Grants No. 12105099.

\end{acknowledgements}

\begin{appendix}
\section{Isospin sum rules for two- and three-body $B$ decays}\label{rules}

\subsection{$b\to c\overline u d/s$ modes}
The isospin sum rules for two- and three-body $B$ decays with $b$ decaying into $c\overline u d/s$, are listed below. 
\begin{align}\label{two1}
{ SumI_-^3}\,[B^-, D^+,\pi^+]&=6\,\big[ \mathcal{A}(B^-\to D^0\pi^-)+\sqrt{2}\,\mathcal{A}(\overline B^0\to D^0\pi^0)-\mathcal{A}(\overline B^0\to D^+\pi^-)\big]=0,
\end{align}
\begin{align}
{ SumI_-^2}\,[B^-, D^+,\overline K^0]&=2\,\big[ \mathcal{A}(B^-\to D^0K^-)-\,\mathcal{A}(\overline B^0\to D^0\overline K^0)-\mathcal{A}(\overline B^0\to D^+K^-)\big]=0,
\end{align}
\begin{align}\label{two2}
{ SumI_-^2}\,[\overline B^0_s, D^+,\pi^+]&=2\,\big[ \sqrt{2}\,\mathcal{A}(\overline B^0_s\to D^0\pi^0)-\,\mathcal{A}(\overline B^0_s\to D^+\pi^-)\big]=0,
\end{align}
\begin{align}
{ SumI_-^4}\,[B^-,D^+,\pi^+,\pi^+]=&\,-24\,\big[\,\sqrt{2}\,\mathcal{A}( B^-\to D^0\pi^0 \pi^-)+\sqrt{2}\,\mathcal{A}( B^-\to D^0\pi^- \pi^0)\nonumber\\&-\mathcal{A}( B^-\to D^+\pi^- \pi^-)+2\,\mathcal{A}( \overline B^0\to D^0\pi^0 \pi^0)\nonumber\\&-\mathcal{A}( \overline B^0\to D^0\pi^- \pi^+)-\mathcal{A}( \overline B^0\to D^0\pi^+ \pi^-)\nonumber\\&-\sqrt{2}\,\mathcal{A}( \overline B^0\to D^+\pi^0 \pi^-)-\sqrt{2}\,\mathcal{A}( \overline B^0\to D^+\pi^- \pi^0)\big]=0,
\end{align}
\begin{align}
{ SumI_-^3}\,[\overline B^0,D^+,\pi^+,\pi^+]=&\,-6\,\big[\,2\,\mathcal{A}( \overline B^0\to D^0\pi^0 \pi^0)-\mathcal{A}( \overline B^0\to D^0\pi^- \pi^+)\nonumber\\&-\mathcal{A}( \overline B^0\to D^0\pi^+ \pi^-)-\sqrt{2}\,\mathcal{A}( \overline B^0\to D^+\pi^0 \pi^-)\nonumber\\&-\sqrt{2}\,\mathcal{A}( \overline B^0\to D^+\pi^- \pi^0)\big]=0,
\end{align}
\begin{align}
{ SumI_-^3}\,[B^-,D^+,\pi^+,\pi^0]=&\,6\,\big[\,2\,\mathcal{A}( B^-\to D^0\pi^0 \pi^-)+\mathcal{A}( B^-\to D^0\pi^- \pi^0)\nonumber\\&-\sqrt{2}\,\mathcal{A}( B^-\to D^+\pi^- \pi^-)+\sqrt{2}\,\mathcal{A}( \overline B^0\to D^0\pi^0 \pi^0)\nonumber\\&-\sqrt{2}\,\mathcal{A}( \overline B^0\to D^0\pi^+ \pi^-)-2\,\mathcal{A}( \overline B^0\to D^+\pi^0 \pi^-)\nonumber\\&-\mathcal{A}( \overline B^0\to D^+\pi^- \pi^0)\big]=0,
\end{align}
\begin{align}
{ SumI_-^3}\,[B^-,D^+,\pi^+,\eta_8]=&\,6\,\big[\,\mathcal{A}( B^-\to D^0\pi^- \eta_8)+\sqrt{2}\,\mathcal{A}( \overline B^0\to D^0\pi^0 \eta_8)\nonumber\\&-\mathcal{A}( \overline B^0\to D^+\pi^- \eta_8)\big]=0,
\end{align}
\begin{align}
{ SumI_-^3}\,[B^-,D^+,K^+,\overline K^0]=&\,6\,\big[\,\mathcal{A}( B^-\to D^0 K^0 K^-)-\mathcal{A}( \overline B^0\to D^0K^0 \overline K^0)\nonumber\\&+\mathcal{A}( \overline B^0\to D^0 K^+ K^-)-\mathcal{A}( \overline B^0\to D^+ K^0 K^-)\big]=0,
\end{align}
\begin{align}
{ SumI_-^3}\,[B^-,D^+_s,\pi^+,\overline K^0]=&\,6\,\big[\,\mathcal{A}( B^-\to D^+_s \pi^-K^-)+\sqrt{2}\,\mathcal{A}( \overline B^0\to D^+_s \pi^0K^-)\nonumber\\&-\mathcal{A}( \overline B^0\to D^+_s \pi^-\overline K^0)\big]=0,
\end{align}
\begin{align}
{ SumI_-^3}\,[\overline B^0_s,D^+,\pi^+, K^+]=&\,6\,\big[\,\sqrt{2}\,\mathcal{A}( \overline B^0_s\to D^0 \pi^0K^0)+\mathcal{A}( \overline B^0_s\to D^0 \pi^-K^+)\nonumber\\&-\mathcal{A}( \overline B^0_s\to D^+ \pi^-K^0)\big]=0,
\end{align}
\begin{align}
{ SumI_-^3}\,[\overline B^0_s,D^+_s,\pi^+, \pi^+]=&\,6\sqrt{2}\,\big[\,\mathcal{A}( \overline B^0_s\to D^+_s \pi^0\pi^-)+\mathcal{A}( \overline B^0_s\to D^+_s \pi^-\pi^0)\big]=0,
\end{align}
\begin{align}
{ SumI_-^3}\,[B^-,D^+,\pi^+,\overline K^0]=&\,6\,\big[-\sqrt{2}\,\mathcal{A}( B^-\to D^0\pi^0 K^-)+\mathcal{A}( B^-\to D^0\pi^- \overline K^0)\nonumber\\&+\mathcal{A}( B^-\to D^+\pi^- K^-)+\sqrt{2}\,\mathcal{A}( \overline B^0\to D^0\pi^0 \overline K^0)\nonumber\\&+\mathcal{A}( \overline B^0\to D^0\pi^+ K^-)+\sqrt{2}\,\mathcal{A}( \overline B^0\to D^+\pi^0 K^-)\nonumber\\&-\mathcal{A}( \overline B^0\to D^+\pi^- \overline K^0)\big]=0,
\end{align}
\begin{align}
{ SumI_-^2}\,[B^-,D^+,\pi^+,\overline K^0]=&\,2\,\big[\,\sqrt{2}\,\mathcal{A}( \overline B^0\to D^0\pi^0 \overline K^0)+\mathcal{A}( \overline B^0\to D^0\pi^+ K^-)\nonumber\\&+\sqrt{2}\,\mathcal{A}( \overline B^0\to D^+\pi^0 K^-)-\mathcal{A}( \overline B^0\to D^+\pi^- \overline K^0)\big]=0,
\end{align}
\begin{align}
{ SumI_-^2}\,[B^-,D^0,\pi^+,\overline K^0]=&\,2\,\big[\sqrt{2}\,\mathcal{A}( B^-\to D^0\pi^0 K^-)-\mathcal{A}( B^-\to D^0\pi^- \overline K^0)\nonumber\\&-\sqrt{2}\,\mathcal{A}( \overline B^0\to D^0\pi^0 \overline K^0)-\mathcal{A}( \overline B^0\to D^0\pi^+ K^-)\big]=0,
\end{align}
\begin{align}
{ SumI_-^2}\,[B^-,D^+,\pi^0,\overline K^0]=&\,2\,\big[\,\mathcal{A}( B^-\to D^0\pi^0 K^-)-\sqrt{2}\,\mathcal{A}( B^-\to D^0\pi^- \overline K^0)\nonumber\\&-\sqrt{2}\,\mathcal{A}( B^-\to D^+\pi^- K^-)-\mathcal{A}( \overline B^0\to D^0\pi^0 \overline K^0)\nonumber\\&-\mathcal{A}( \overline B^0\to D^+\pi^0 K^-)+\sqrt{2}\,\mathcal{A}( \overline B^0\to D^+\pi^- \overline K^0)\big]=0,
\end{align}
\begin{align}
{ SumI_-^2}\,[B^-,D^+,\pi^0,K^-]=&\,2\,\big[\,\sqrt{2}\,\mathcal{A}( B^-\to D^0\pi^0 K^-)-\mathcal{A}( B^-\to D^+\pi^- K^-)\nonumber\\&-\mathcal{A}( \overline B^0\to D^0\pi^+ K^-)-\sqrt{2}\,\mathcal{A}( \overline B^0\to D^+\pi^0 K^-)\big]=0,
\end{align}
\begin{align}
{ SumI_-^2}\,[B^-,D^+,\overline K^0,\eta_8]=&\,2\,\big[\,\mathcal{A}( B^-\to D^0 K^-\eta_8)-\mathcal{A}( \overline B^0\to D^0 \overline K^0\eta_8)\nonumber\\&-\mathcal{A}( \overline B^0\to D^+K^-\eta_8)\big]=0,
\end{align}
\begin{align}
{ SumI_-^2}\,[B^-,D^+_s,\overline K^0,\overline K^0]=&\,2\,\big[\,\mathcal{A}( B^-\to D^+_s K^-K^-)-\mathcal{A}( \overline B^0\to D^+_s K^-\overline K^0)\nonumber\\&-\mathcal{A}( \overline B^0\to D^+_s\overline K^0K^-)\big]=0,
\end{align}
\begin{align}
{ SumI_-^3}\,[\overline B^0_s,D^+,\pi^+,\pi^+]=&\,6\,\big[-2\,\mathcal{A}( \overline B^0_s\to D^0\pi^0 \pi^0)+\mathcal{A}(\overline B^0_s\to D^0\pi^- \pi^+)\nonumber\\&+\mathcal{A}(\overline B^0_s\to D^0\pi^+ \pi^-)+\sqrt{2}\,\mathcal{A}( \overline B^0_s\to D^+\pi^0 \pi^-)\nonumber\\&+\sqrt{2}\,\mathcal{A}( \overline B^0_s\to D^+\pi^- \pi^0)\big]=0,
\end{align}
\begin{align}
{ SumI_-^2}\,[\overline B^0_s,D^0,\pi^+,\pi^+]=&\,2\,\big[\,2\,\mathcal{A}( \overline B^0_s\to D^0\pi^0 \pi^0)-\mathcal{A}(\overline B^0_s\to D^0\pi^- \pi^+)\nonumber\\&-\mathcal{A}(\overline B^0_s\to D^0\pi^+ \pi^-)\big]=0,
\end{align}
\begin{align}
{ SumI_-^2}\,[\overline B^0_s,D^+,\pi^+,\pi^0]=&\,2\,\big[\,\sqrt{2}\,\mathcal{A}( \overline B^0_s\to D^0\pi^0 \pi^0)-\sqrt{2}\,\mathcal{A}(\overline B^0_s\to D^0\pi^+ \pi^-)\nonumber\\&-2\,\mathcal{A}( \overline B^0_s\to D^+\pi^0 \pi^-)-\mathcal{A}( \overline B^0_s\to D^+\pi^- \pi^0)\big]=0,
\end{align}
\begin{align}
{ SumI_-^2}\,[\overline B^0_s,D^+,\pi^+,\eta_8]=&\,2\,\big[\,\sqrt{2}\,\mathcal{A}( \overline B^0_s\to D^0\pi^0 \eta_8)-\mathcal{A}( \overline B^0_s\to D^+\pi^- \eta_8)\big]=0,
\end{align}
\begin{align}
{ SumI_-^2}\,[\overline B^0_s,D^+,K^+,\overline K^0]=&-2\,\big[\,\mathcal{A}( \overline B^0_s\to D^0K^0\overline K^0)-\mathcal{A}( \overline B^0_s\to D^0 K^+ K^-)\nonumber\\&+\mathcal{A}( \overline B^0_s\to D^+ K^0 K^-)\big]=0,
\end{align}
\begin{align}
{ SumI_-^2}\,[\overline B^0_s,D^+_s,\pi^+,\overline K^0]=&\,2\,\big[\,\sqrt{2}\,\mathcal{A}( \overline B^0_s\to D^+_s\pi^0 K^-)-\mathcal{A}( \overline B^0_s\to D^+_s\pi^- \overline K^0)\big]=0.
\end{align}

\subsection{$b\to c\overline c d/s$ modes}\label{cc}
The isospin sum rules for two- and three-body $B$ decays with $b$ decaying into $c\overline c d/s$, are listed below.
\begin{align}\label{cc1}
{ SumI_-^3}\,[B^-,D^+,\overline{D}^0,\pi ^+]=&6\,\big[\,\sqrt{2}\,\mathcal{A}(B^-\to D^0D^-\pi^0)+\mathcal{A}(B^-\to D^0
\overline D^0\pi^-)\nonumber\\&-\mathcal{A}(B^-\to D^+
 D^-\pi^-)-\mathcal{A}(\overline B^0\to D^0
 D^-\pi^+)\nonumber\\&+\sqrt{2}\,\mathcal{A}(\overline B^0\to D^0\overline D^0\pi^0)-\sqrt{2}\,\mathcal{A}(\overline B^0\to D^+D^-\pi^0)\nonumber\\&-\mathcal{A}(\overline B^0\to D^+\overline D^0\pi^-)\big]=0,
\end{align}
\begin{align}\label{r1}
{ SumI_-^2}\,[B^-,D^+,\overline{D}^0,\overline K^0]=&-2\,\big[\,\mathcal{A}(B^-\to D^0D^-\overline K^0)-\mathcal{A}(B^-\to D^0
\overline D^0K^-)\nonumber\\&+\mathcal{A}(B^-\to D^+
 D^-K^-)+\mathcal{A}(\overline B^0\to D^0
 \overline D^0\overline K^0)\nonumber\\&-\mathcal{A}(\overline B^0\to D^+D^-\overline K^0)+\mathcal{A}(\overline B^0\to D^+\overline D^0 K^-)\big]=0,
\end{align}
\begin{align}
{ SumI_-}\,[\overline B^0,D^+,\overline{D}^0,\overline K^0]=&-\mathcal{A}(\overline B^0\to D^0\overline D^0\overline K^0)+\mathcal{A}(\overline B^0\to D^+D^-\overline K^0)\nonumber\\&-\mathcal{A}(\overline B^0\to D^+\overline D^0 K^-)=0,
\end{align}
\begin{align}
{ SumI_-}\,[B^-,D^0,\overline{D}^0,\overline K^0]=&\,\mathcal{A}(B^-\to D^0D^-\overline K^0)-\mathcal{A}(B^-\to D^0
\overline D^0K^-)\nonumber\\&+\mathcal{A}(\overline B^0\to D^0
 \overline D^0\overline K^0)=0,
\end{align}
\begin{align}
{ SumI_-}\,[B^-,D^+,D^-,\overline K^0]=&-\mathcal{A}(B^-\to D^0D^-\overline K^0)-\mathcal{A}(B^-\to D^+
D^-K^-)\nonumber\\&+\mathcal{A}(\overline B^0\to D^+
D^-\overline K^0)=0,
\end{align}
\begin{align}\label{r2}
{ SumI_-}\,[B^-,D^+,\overline D^0, K^-]=&-\mathcal{A}(B^-\to D^0\overline D^0K^-)+\mathcal{A}(B^-\to D^+
D^-K^-)\nonumber\\&+\mathcal{A}(\overline B^0\to D^+
\overline D^0 K^-)=0,
\end{align}
\begin{align}\label{r3}
{ SumI_-^2}\,[B^-,D^+,D^-_s,\pi^+]=&\,2\,\big[\,\sqrt{2}\,\mathcal{A}(B^-\to D^0D^-_s \pi^0)-\mathcal{A}(B^-\to D^+D^-_s \pi^-)\nonumber\\&-\mathcal{A}(\overline B^0\to D^0D^-_s \pi^+)-\sqrt{2}\,\mathcal{A}(\overline B^0\to D^+D^-_s \pi^0)\big]=0,
\end{align}
\begin{align}
{ SumI_-}\,[\overline B^0,D^+,D^-_s,\pi^+]=&-\mathcal{A}(\overline B^0\to D^0D^-_s \pi^+)-\sqrt{2}\,\mathcal{A}(\overline B^0\to D^+D^-_s \pi^0)=0,
\end{align}
\begin{align}
{ SumI_-}\,[ B^-,D^0,D^-_s,\pi^+]=&-\sqrt{2}\,\mathcal{A}(B^-\to D^0D^-_s \pi^0)+\mathcal{A}(\overline B^0\to D^0D^-_s \pi^+)=0,
\end{align}
\begin{align}\label{r4}
{ SumI_-}\,[ B^-,D^+,D^-_s,\pi^0]=&-\mathcal{A}(B^-\to D^0D^-_s \pi^0)+\sqrt{2}\,\mathcal{A}(B^-\to D^+D^-_s \pi^-)\nonumber\\&+\mathcal{A}(\overline B^0\to D^+D^-_s \pi^0)=0,
\end{align}
\begin{align}
{ SumI_-}\,[ B^-,D^+,D^-_s,\eta_8]=&-\mathcal{A}(B^-\to D^0D^-_s \eta_8)+\mathcal{A}(\overline B^0\to D^+D^-_s \eta_8)=0,
\end{align}
\begin{align}
{ SumI_-}\,[ B^-,D^+_s,D^-_s,\overline K^0]=&-\mathcal{A}(B^-\to D^+_sD^-_s K^-)+\mathcal{A}(\overline B^0\to D^+_sD^-_s \overline K^0)=0,
\end{align}
\begin{align}\label{cc2}
{ SumI_-^2}\,[ \overline B^0_s,D^+,\overline D^0,\pi^+]=&-2\,\big[\mathcal{A}(\overline B^0_s\to D^0D^- \pi^+)-\sqrt{2}\,\mathcal{A}(\overline B^0_s\to D^0\overline D^0 \pi^0)\nonumber\\&+\sqrt{2}\,\mathcal{A}(\overline B^0_s\to D^+D^- \pi^0)+\mathcal{A}(\overline B^0_s\to D^+\overline D^0 \pi^-)\big]=0,
\end{align}
\begin{align}
{ SumI_-}\,[ \overline B^0_s,D^0,\overline D^0,\pi^+]=&\mathcal{A}(\overline B^0_s\to D^0D^- \pi^+)-\sqrt{2}\,\mathcal{A}(\overline B^0_s\to D^0\overline D^0 \pi^0)=0,
\end{align}
\begin{align}
{ SumI_-}\,[ \overline B^0_s,D^+,D^-,\pi^+]=&-\mathcal{A}(\overline B^0_s\to D^0D^- \pi^+)-\sqrt{2}\,\mathcal{A}(\overline B^0_s\to D^+ D^- \pi^0)=0,
\end{align}
\begin{align}
{ SumI_-}\,[ \overline B^0_s,D^+,\overline D^0,\pi^0]=&-\mathcal{A}(\overline B^0_s\to D^0\overline D^0 \pi^0)+\mathcal{A}(\overline B^0_s\to D^+D^- \pi^0)\nonumber\\&+\sqrt{2}\,\mathcal{A}(\overline B^0_s\to D^+\overline D^0 \pi^-)=0,
\end{align}
\begin{align}
{ SumI_-}\,[ \overline B^0_s,D^+,\overline D^0,\eta_8]=&-\mathcal{A}(\overline B^0_s\to D^0\overline D^0 \eta_8)+\mathcal{A}(\overline B^0_s\to D^+D^- \eta_8)=0,
\end{align}
\begin{align}
{ SumI_-}\,[ \overline B^0_s,D^+,D^-_s,K^+]=&-\mathcal{A}(\overline B^0_s\to D^0D^-_s K^+)+\mathcal{A}(\overline B^0_s\to D^+D^-_s K^0)=0,
\end{align}
\begin{align}
{ SumI_-}\,[ \overline B^0_s,D^+_s,\overline D^0,\overline K^0]=&\,\mathcal{A}(\overline B^0_s\to D^+_sD^- \overline K^0)-\mathcal{A}(\overline B^0_s\to D^+_s\overline D^0 K^-)=0,
\end{align}
\begin{align}
{ SumI_-}\,[ \overline B^0_s,D^+_s,D^-_s,\pi^+]=&-\sqrt{2}\,\mathcal{A}(\overline B^0_s\to D^+_sD^-_s \pi^0)=0,
\end{align}
\begin{align}\label{cc3}
{ SumI_-^3}\,[ B^-,J/\Psi,\pi^+,\pi^+]=&6\,\big[\,\sqrt{2}\,\mathcal{A}( B^-\to J/\Psi\pi^0 \pi^-)+\sqrt{2}\,\mathcal{A}( B^-\to J/\Psi\pi^- \pi^0)\nonumber\\&+2\,\mathcal{A}(\overline B^0\to J/\Psi\pi^0 \pi^0)-\mathcal{A}(\overline B^0\to J/\Psi\pi^- \pi^+)\nonumber\\&-\mathcal{A}(\overline B^0\to J/\Psi\pi^+ \pi^-)\big]=0,
\end{align}
\begin{align}\label{r5}
{ SumI_-^2}\,[ B^-,J/\Psi,\pi^+,\overline K^0]=&2\,\big[\,\sqrt{2}\,\mathcal{A}( B^-\to J/\Psi\pi^0 K^-)-\mathcal{A}( B^-\to J/\Psi\pi^- \overline K^0)\nonumber\\&-\sqrt{2}\,\mathcal{A}(\overline B^0\to J/\Psi\pi^0 \overline K^0)-\mathcal{A}(\overline B^0\to J/\Psi\pi^+ K^-)\big]=0,
\end{align}
\begin{align}
{ SumI_-}\,[\overline B^0,J/\Psi,\pi^+,\overline K^0]=&-\sqrt{2}\,\mathcal{A}(\overline B^0\to J/\Psi\pi^0 \overline K^0)-\mathcal{A}(\overline B^0\to J/\Psi\pi^+ K^-)=0,
\end{align}
\begin{align}
{ SumI_-}\,[ B^-,J/\Psi,\pi^0,\overline K^0]=&-\mathcal{A}( B^-\to J/\Psi\pi^0 K^-)+\sqrt{2}\,\mathcal{A}( B^-\to J/\Psi\pi^- \overline K^0)\nonumber\\&+\mathcal{A}(\overline B^0\to J/\Psi\pi^0 \overline K^0)=0,
\end{align}
\begin{align}\label{r6}
{ SumI_-}\,[ B^-,J/\Psi,\pi^0,K^-]=&-\sqrt{2}\,\mathcal{A}( B^-\to J/\Psi\pi^0 K^-)+\mathcal{A}( \overline B^0\to J/\Psi\pi^+  K^-)=0,
\end{align}
\begin{align}
{ SumI_-}\,[ B^-,J/\Psi,\overline K^0,\eta_8]=&-\mathcal{A}( B^-\to J/\Psi K^-\eta_8)+\mathcal{A}(\overline B^0\to J/\Psi \overline K^0\eta_8)=0,
\end{align}
\begin{align}
{ SumI_-}\,[ \overline B^0_s,J/\Psi,\pi^+,\pi^+]=&\,2\,\big[\,2\,\mathcal{A}( \overline B^0_s\to J/\Psi\pi^0 \pi^0)-\mathcal{A}( \overline B^0_s\to J/\Psi\pi^- \pi^+)\nonumber\\&-\mathcal{A}(\overline B^0_s\to J/\Psi\pi^+ \pi^-)\big]=0,
\end{align}
\begin{align}
{ SumI_-}\,[ \overline B^0_s,J/\Psi,\pi^+,\pi^0]=&-\sqrt{2}\,\mathcal{A}( \overline B^0_s\to J/\Psi\pi^0 \pi^0)+\sqrt{2}\,\mathcal{A}( \overline B^0_s\to J/\Psi\pi^+ \pi^-)=0,
\end{align}
\begin{align}
{ SumI_-}\,[ \overline B^0_s,J/\Psi,\pi^+,\eta_8]=&-\sqrt{2}\,\mathcal{A}( \overline B^0_s\to J/\Psi\pi^0 \eta_8)=0,
\end{align}
\begin{align}
{ SumI_-}\,[ \overline B^0_s,J/\Psi,K^+,\overline K^0]=&\,\mathcal{A}( \overline B^0_s\to J/\Psi K^0 \overline K^0)-\mathcal{A}( \overline B^0_s\to J/\Psi K^+ K^-)=0.
\end{align}

\subsection{$b\to u\overline u d/s$ modes}\label{uu}
The isospin sum rules for two- and three-body $B$ decays with $b$ decaying into $u\overline u d/s$, are listed below.
\begin{align}\label{two3}
{ SumI_-^3}\,[B^-, \pi^+,\pi^+]&=12\,\big[\sqrt{2}\,\mathcal{A}(B^-\to \pi^0\pi^-)+\,\mathcal{A}(\overline B^0\to \pi^0\pi^0)-\mathcal{A}(\overline B^0\to \pi^+\pi^-)\big]=0,
\end{align}
\begin{align}
{ SumI_-^2}\,[B^-, \pi^+,\overline K^0]&=2\,\big[ \sqrt{2}\,\mathcal{A}(B^-\to \pi^0K^-)-\mathcal{A}(B^-\to \pi^-\overline K^0)\nonumber\\&~~~~~-\sqrt{2}\,\mathcal{A}(\overline B^0\to \pi^0\overline K^0)-\mathcal{A}(\overline B^0\to \pi^+K^-)\big]=0,
\end{align}
\begin{align}\label{two4}
{ SumI_-^2}\,[\overline B^0_s, \pi^+,\pi^+]&=4\,\big[\, \mathcal{A}(\overline B^0_s\to \pi^0\pi^0)-\,\mathcal{A}(\overline B^0_s\to \pi^+\pi^-)\big]=0,
\end{align}
\begin{align}
{ SumI_-^4}\,[B^-,\pi^+,\pi^+,\pi^+]=&\,24\,\big[-2\,\mathcal{A}( B^-\to \pi^0\pi^0 \pi^-)-2\,\mathcal{A}( B^-\to \pi^0\pi^- \pi^0)\nonumber\\&-2\,\mathcal{A}( B^-\to \pi^-\pi^0 \pi^0)+\mathcal{A}( B^-\to \pi^-\pi^- \pi^+)+\mathcal{A}( B^-\to \pi^-\pi^+ \pi^-)\nonumber\\&+\mathcal{A}( B^-\to \pi^+\pi^- \pi^-)-2\sqrt{2}\,\mathcal{A}( \overline B^0\to  \pi^0 \pi^0 \pi^0)+\sqrt{2}\,\mathcal{A}( \overline B^0\to  \pi^0 \pi^- \pi^+)\nonumber\\&+\sqrt{2}\,\mathcal{A}( \overline B^0\to \pi^0 \pi^+ \pi^-)+\sqrt{2}\,\mathcal{A}( \overline B^0\to  \pi^-\pi^0 \pi^+)\nonumber\\&+\sqrt{2}\,\mathcal{A}( \overline B^0\to  \pi^+\pi^0 \pi^-)+\sqrt{2}\,\mathcal{A}( \overline B^0\to \pi^-\pi^+ \pi^0)\nonumber\\&+\sqrt{2}\,\mathcal{A}( \overline B^0\to \pi^+ \pi^-\pi^0)\big]=0,
\end{align}
\begin{align}
{ SumI_-^3}\,[B^-,\pi^+,\pi^+,\pi^+]=&\,6\sqrt{2}\,\big[-2\,\mathcal{A}( \overline B^0\to  \pi^0 \pi^0 \pi^0)+\mathcal{A}( \overline B^0\to  \pi^0 \pi^- \pi^+)\nonumber\\&+\mathcal{A}( \overline B^0\to \pi^0 \pi^+ \pi^-)+\mathcal{A}( \overline B^0\to  \pi^-\pi^0 \pi^+)+\mathcal{A}( \overline B^0\to  \pi^+\pi^0 \pi^-)\nonumber\\&+\mathcal{A}( \overline B^0\to \pi^-\pi^+ \pi^0)+\mathcal{A}( \overline B^0\to \pi^+ \pi^-\pi^0)\big]=0,
\end{align}
\begin{align}
{ SumI_-^3}\,[B^-,\pi^+,\pi^+,\pi^0]=&\,6\,\big[\,2\sqrt{2}\,\mathcal{A}( B^-\to \pi^0\pi^0 \pi^-)+\sqrt{2}\,\mathcal{A}( B^-\to \pi^0\pi^- \pi^0)\nonumber\\&+\sqrt{2}\,\mathcal{A}( B^-\to \pi^-\pi^0 \pi^0)-\sqrt{2}\,\mathcal{A}( B^-\to \pi^-\pi^+ \pi^-)\nonumber\\&-\sqrt{2}\,\mathcal{A}( B^-\to \pi^+\pi^- \pi^-)+2\,\mathcal{A}( \overline B^0\to  \pi^0 \pi^0 \pi^0)\nonumber\\&-2\,\mathcal{A}( \overline B^0\to  \pi^0 \pi^+ \pi^-)-2\,\mathcal{A}( \overline B^0\to \pi^+ \pi^0 \pi^-)\nonumber\\&-\mathcal{A}( \overline B^0\to  \pi^-\pi^+ \pi^0)-\mathcal{A}( \overline B^0\to  \pi^+\pi^- \pi^0)\big]=0,
\end{align}
\begin{align}
{ SumI_-^3}\,[B^-,\pi^+,\pi^+,\eta_8]=&\,6\,\big[\,\sqrt{2}\,\mathcal{A}( B^-\to \pi^0\pi^- \eta_8)+\sqrt{2}\,\mathcal{A}( B^-\to \pi^-\pi^0 \eta_8)\nonumber\\&+2\,\mathcal{A}( \overline B^0\to  \pi^0 \pi^0 \eta_8)-\mathcal{A}( \overline B^0\to  \pi^-\pi^+ \eta_8)\nonumber\\&-\mathcal{A}( \overline B^0\to  \pi^+\pi^- \eta_8)\big]=0,
\end{align}
\begin{align}
{ SumI_-^3}\,[B^-,\pi^+,K^+,\overline K^0]=&\,6\,\big[\,\sqrt{2}\,\mathcal{A}( B^-\to \pi^0 K^0 K^-)-\mathcal{A}( B^-\to \pi^-K^0 \overline K^0)\nonumber\\&+\mathcal{A}( B^-\to \pi^-K^+K^-)-\sqrt{2}\,\mathcal{A}( \overline B^0\to  \pi^0 K^0 \overline K^0)\nonumber\\&+\sqrt{2}\,\mathcal{A}( \overline B^0\to  \pi^0 K^+ K^-)-\mathcal{A}( \overline B^0\to \pi^+ K^0 K^-)\nonumber\\&-\mathcal{A}( \overline B^0\to  \pi^-K^+ \overline K^0)\big]=0,
\end{align}
\begin{align}
{ SumI_-^3}\,[\overline B^0_s,\pi^+,\pi^+, K^+]=&\,6\,\big[\,2\,\mathcal{A}( \overline B^0_s\to \pi^0 \pi^0 K^0)+\sqrt{2}\,\mathcal{A}( \overline B^0_s\to \pi^0 \pi^- K^+)\nonumber\\&+\sqrt{2}\,\mathcal{A}(\overline B^0_s\to \pi^- \pi^0 K^+)-\mathcal{A}( \overline B^0_s\to \pi^- \pi^+ K^0)\nonumber\\&-\mathcal{A}( \overline B^0_s\to \pi^+ \pi^- K^0)\big]=0,
\end{align}
\begin{align}
{ SumI_-^3}\,[B^-,\pi^+,\pi^+,\overline K^0]=&\,6\,\big[-2\,\mathcal{A}( B^-\to \pi^0 \pi^0 K^-)+\sqrt{2}\,\mathcal{A}( B^-\to \pi^0\pi^- \overline K^0)\nonumber\\&+\sqrt{2}\,\mathcal{A}( B^-\to \pi^-\pi^0\overline K^0)+\mathcal{A}(  B^-\to  \pi^- \pi^+ K^-)\nonumber\\&+\mathcal{A}( B^-\to  \pi^+ \pi^- K^-)+2\,\mathcal{A}( \overline B^0\to \pi^0 \pi^0 \overline K^0)\nonumber\\&+\sqrt{2}\,\mathcal{A}( \overline B^0\to  \pi^0\pi^+ K^-)+\sqrt{2}\,\mathcal{A}( \overline B^0\to  \pi^+\pi^0 K^-)\nonumber\\&-\mathcal{A}( \overline B^0\to  \pi^-\pi^+ \overline K^0)-\mathcal{A}( \overline B^0\to  \pi^+\pi^- \overline K^0)\big]=0,
\end{align}
\begin{align}
{ SumI_-^2}\,[\overline B^0,\pi^+,\pi^+,\overline K^0]=&\,2\,\big[\,2\,\mathcal{A}( \overline B^0\to \pi^0 \pi^0 \overline K^0)+\sqrt{2}\,\mathcal{A}( \overline B^0\to  \pi^0\pi^+ K^-)\nonumber\\&+\sqrt{2}\,\mathcal{A}( \overline B^0\to  \pi^+\pi^0 K^-)-\mathcal{A}( \overline B^0\to  \pi^-\pi^+ \overline K^0)\nonumber\\&-\mathcal{A}( \overline B^0\to  \pi^+\pi^- \overline K^0)\big]=0,
\end{align}
\begin{align}
{ SumI_-^2}\,[B^-,\pi^+,\pi^0,\overline K^0]=&\,2\,\big[\,\sqrt{2}\,\mathcal{A}( B^-\to \pi^0 \pi^0 K^-)-2\,\mathcal{A}( B^-\to \pi^0\pi^- \overline K^0)\nonumber\\&-\mathcal{A}( B^-\to \pi^-\pi^0\overline K^0)-\sqrt{2}\,\mathcal{A}(  B^-\to  \pi^+ \pi^- K^-)\nonumber\\&-\sqrt{2}\,\mathcal{A}( \overline B^0\to \pi^0 \pi^0 \overline K^0)-\mathcal{A}( \overline B^0\to  \pi^+\pi^0 K^-)\nonumber\\&+\sqrt{2}\,\mathcal{A}( \overline B^0\to  \pi^+\pi^- \overline K^0)\big]=0,
\end{align}
\begin{align}
{ SumI_-^2}\,[B^-,\pi^+,\pi^+,K^-]=&\,2\,\big[\,2\,\mathcal{A}( B^-\to \pi^0 \pi^0 K^-)-\mathcal{A}(  B^-\to  \pi^- \pi^+ K^-)\nonumber\\&-\mathcal{A}( B^-\to  \pi^+ \pi^- K^-)-\sqrt{2}\,\mathcal{A}( \overline B^0\to  \pi^0\pi^+ K^-)\nonumber\\&-\sqrt{2}\,\mathcal{A}( \overline B^0\to  \pi^+\pi^0 K^-)\big]=0,
\end{align}
\begin{align}
{ SumI_-^2}\,[B^-,\pi^+,\overline K^0,\eta_8]=&\,2\,\big[\,\sqrt{2}\,\mathcal{A}( B^-\to \pi^0 K^-\eta_8)-\mathcal{A}( B^-\to \pi^-\overline K^0\eta_8)\nonumber\\&-\sqrt{2}\,\mathcal{A}( \overline B^0\to \pi^0 \overline K^0\eta_8)-\mathcal{A}( \overline B^0\to  \pi^+ K^-\eta_8)\big]=0,
\end{align}
\begin{align}
{ SumI_-^2}\,[B^-,K^+,\overline K^0,\overline K^0]=&-2\,\big[\,\mathcal{A}( B^-\to K^0 K^-\overline K^0)+\mathcal{A}( B^-\to K^0\overline K^0 K^-)\nonumber\\&-\mathcal{A}( B^-\to K^+K^- K^-)-\mathcal{A}( \overline B^0\to  K^0\overline K^0 \overline K^0)\nonumber\\&+\mathcal{A}(  \overline B^0\to K^+K^- \overline K^0)+\mathcal{A}( \overline B^0\to  K^+\overline K^0 K^-)\big]=0,
\end{align}
\begin{align}
{ SumI_-^3}\,[\overline B^0_s,\pi^+,\pi^+,\pi^+]=&\,6\sqrt{2}\,\big[-2\,\mathcal{A}( \overline B^0_s\to \pi^0 \pi^0\pi^0)+\mathcal{A}(  \overline B^0_s\to \pi^0 \pi^-\pi^+)\nonumber\\&+\mathcal{A}(\overline B^0_s\to \pi^0 \pi^+\pi^-)+\mathcal{A}(\overline B^0_s\to \pi^- \pi^0\pi^+)+\mathcal{A}(\overline B^0_s\to \pi^+ \pi^0\pi^-)\nonumber\\&+\mathcal{A}(\overline B^0_s\to \pi^- \pi^+\pi^0)+\mathcal{A}(\overline B^0_s\to \pi^+ \pi^-\pi^0)\big]=0,
\end{align}
\begin{align}
{ SumI_-^2}\,[\overline B^0_s,\pi^+,\pi^+,\pi^0]=&\,2\,\big[\,2\,\mathcal{A}( \overline B^0_s\to \pi^0 \pi^0\pi^0)-2\,\mathcal{A}(\overline B^0_s\to \pi^0 \pi^+\pi^-)-2\,\mathcal{A}(\overline B^0_s\to \pi^+ \pi^0\pi^-)\nonumber\\&-\mathcal{A}(\overline B^0_s\to \pi^- \pi^+\pi^0)-\mathcal{A}(\overline B^0_s\to \pi^+ \pi^-\pi^0)\big]=0,
\end{align}
\begin{align}
{ SumI_-^2}\,[\overline B^0_s,\pi^+,\pi^+,\eta_8]=&\,2\,\big[\,2\,\mathcal{A}( \overline B^0_s\to \pi^0 \pi^0\eta_8)-\mathcal{A}(\overline B^0_s\to \pi^- \pi^+\eta_8)\nonumber\\&-\mathcal{A}(\overline B^0_s\to \pi^+ \pi^-\eta_8)\big]=0,
\end{align}
\begin{align}
{ SumI_-^2}\,[\overline B^0_s,\pi^+,K^+,\overline K^0]=&-2\,\big[\,\sqrt{2}\,\mathcal{A}( \overline B^0_s\to \pi^0 K^0\overline K^0)-\sqrt{2}\,\mathcal{A}(\overline B^0_s\to \pi^0 K^+K^-)\nonumber\\&+\mathcal{A}( \overline B^0_s\to \pi^+ K^0 K^-)+\mathcal{A}(\overline B^0_s\to \pi^- K^+\overline K^0)\big]=0.
\end{align}

\subsection{$b\to u\overline c d/s$ modes}
The isospin sum rules for two- and three-body $B$ decays with $b$ decaying into $u\overline c d/s$, are listed below.
\begin{align}
{ SumI_-^2}\,[B^-,\overline D^0,\pi^+]=&-2\,\big[\,\sqrt{2}\,\mathcal{A}( B^-\to D^-\pi^0)+\mathcal{A}( B^-\to \overline D^0 \pi^-)\nonumber\\&-\mathcal{A}( \overline B^0\to  D^-\pi^+)+\sqrt{2}\,\mathcal{A}( \overline B^0\to  \overline D^0\pi^0)\big]=0,
\end{align}
\begin{align}
{ SumI_-}\,[B^-,\overline D^0,\overline K^0]=&\,\mathcal{A}( B^-\to D^-\overline K^0)-\mathcal{A}( B^-\to \overline D^0 K^-)\nonumber\\&+\mathcal{A}( \overline B^0\to  \overline D^0\overline K^0)\big]=0,
\end{align}
\begin{align}
{ SumI_-}\,[B^-, D^-_s,\pi^+]=&-\sqrt{2}\,\mathcal{A}( B^-\to D^-_s\pi^0)+\mathcal{A}( \overline B^0\to D^-_s\pi^+)\big]=0,
\end{align}
\begin{align}
{ SumI_-}\,[\overline B^0_s,\overline D^0,\pi^+]=&\,\mathcal{A}( \overline B^0_s\to  D^-\pi^+)-\sqrt{2}\,\mathcal{A}( \overline B^0_s\to  \overline D^0\pi^0)=0,
\end{align}
\begin{align}
{ SumI_-^3}\,[B^-,\overline D^0,\pi^+,\pi^+]=&-6\,\big[-2\,\mathcal{A}( B^-\to D^-\pi^0 \pi^0)+\mathcal{A}( B^-\to D^-\pi^- \pi^+)\nonumber\\&+\mathcal{A}( B^-\to D^-\pi^+ \pi^-)-\sqrt{2}\,\mathcal{A}( B^-\to \overline D^0\pi^0 \pi^-)\nonumber\\&-\sqrt{2}\,\mathcal{A}( B^-\to \overline D^0\pi^- \pi^0)+\sqrt{2}\,\mathcal{A}( \overline B^0\to  D^-\pi^0 \pi^+)\nonumber\\&+\sqrt{2}\,\mathcal{A}( \overline B^0\to D^-\pi^+ \pi^0)-2\,\mathcal{A}( \overline B^0\to  \overline D^0\pi^0 \pi^0)\nonumber\\&+\mathcal{A}( \overline B^0\to \overline D^0\pi^- \pi^+)+\mathcal{A}( \overline B^0\to  \overline D^0\pi^+ \pi^-)\big]=0,
\end{align}
\begin{align}
{ SumI_-^2}\,[\overline B^0,\overline D^0,\pi^+,\pi^+]=&-2\,\big[\,\sqrt{2}\,\mathcal{A}( \overline B^0\to  D^-\pi^0 \pi^+)+\sqrt{2}\,\mathcal{A}( \overline B^0\to D^-\pi^+ \pi^0)\nonumber\\&-2\,\mathcal{A}( \overline B^0\to  \overline D^0\pi^0 \pi^0)+\mathcal{A}( \overline B^0\to \overline D^0\pi^- \pi^+)\nonumber\\&+\mathcal{A}( \overline B^0\to  \overline D^0\pi^+ \pi^-)\big]=0,
\end{align}
\begin{align}
{ SumI_-^2}\,[B^-,D^-,\pi^+,\pi^+]=&-2\,\big[-2\,\mathcal{A}( B^-\to D^-\pi^0 \pi^0)+\mathcal{A}( B^-\to D^-\pi^- \pi^+)\nonumber\\&+\mathcal{A}( B^-\to D^-\pi^+ \pi^-)+\sqrt{2}\,\mathcal{A}( \overline B^0\to  D^-\pi^0 \pi^+)\nonumber\\&+\sqrt{2}\,\mathcal{A}( \overline B^0\to D^-\pi^+ \pi^0)\big]=0,
\end{align}
\begin{align}
{ SumI_-^2}\,[B^-,\overline D^0,\pi^+,\pi^0]=&-2\,\big[\,\sqrt{2}\,\mathcal{A}( B^-\to D^-\pi^0 \pi^0)-\sqrt{2}\,\mathcal{A}( B^-\to D^-\pi^+ \pi^-)\nonumber\\&+2\,\mathcal{A}( B^-\to \overline D^0\pi^0 \pi^-)+\mathcal{A}( B^-\to \overline D^0\pi^- \pi^0)-\mathcal{A}( \overline B^0\to  D^-\pi^+ \pi^0)\nonumber\\&+\sqrt{2}\,\mathcal{A}( \overline B^0\to  \overline D^0\pi^0 \pi^0)-\sqrt{2}\,\mathcal{A}( \overline B^0\to \overline D^0\pi^+ \pi^-)\big]=0,
\end{align}
\begin{align}
{ SumI_-^2}\,[B^-,\overline D^0,\pi^+,\eta_8]=&-2\,\big[\,\sqrt{2}\,\mathcal{A}( B^-\to D^-\pi^0 \eta_8)+\mathcal{A}( B^-\to \overline D^0\pi^- \eta_8)\nonumber\\&-\mathcal{A}( \overline B^0\to  D^-\pi^+ \eta_8)+\sqrt{2}\,\mathcal{A}( \overline B^0\to  \overline D^0\pi^0 \eta_8)\big]=0,
\end{align}
\begin{align}
{ SumI_-^2}\,[B^-,\overline D^0,K^+,\overline K^0]=&\,2\,\big[\,\mathcal{A}( B^-\to D^-K^0 \overline K^0)-\mathcal{A}( B^-\to D^-K^+ K^-)\nonumber\\&-\mathcal{A}( B^-\to \overline D^0K^0  K^-)+\mathcal{A}( \overline B^0\to  D^-K^+ \overline K^0)\nonumber\\&+\mathcal{A}( \overline B^0\to \overline D^0K^0  \overline K^0)-\mathcal{A}( \overline B^0\to  \overline D^0K^+  K^-)\big]=0,
\end{align}
\begin{align}
{ SumI_-^2}\,[B^-, D^-_s,\pi^+,K^+]=&-2\,\big[\,\sqrt{2}\,\mathcal{A}( B^-\to D^-_s\pi^0 K^0)+\mathcal{A}( B^-\to D^-_s\pi^- K^+)\nonumber\\&+\sqrt{2}\,\mathcal{A}( \overline B^0\to D^-_s\pi^0 K^+)-\mathcal{A}( \overline B^0\to D^-_s\pi^+ K^0)\big]=0,
\end{align}
\begin{align}
{ SumI_-^2}\,[\overline B^0_s, \overline D^0,\pi^+,K^+]=&-2\,\big[\,\sqrt{2}\,\mathcal{A}( \overline B^0_s\to D^-\pi^0 K^+)-\mathcal{A}( \overline B^0_s\to D^-\pi^+ K^0)\nonumber\\&+\sqrt{2}\,\mathcal{A}( \overline B^0_s\to \overline D^0\pi^0 K^0)+\mathcal{A}( \overline B^0_s\to \overline D^0\pi^- K^+)\big]=0,
\end{align}
\begin{align}
{ SumI_-^2}\,[B^-,\overline D^0,\pi^+,\overline K^0]=&-2\,\big[\,\sqrt{2}\,\mathcal{A}( B^-\to D^-\pi^0 \overline K^0)+\mathcal{A}( B^-\to D^-\pi^+ K^-)\nonumber\\&-\sqrt{2}\,\mathcal{A}( B^-\to \overline D^0\pi^0 K^-)+\mathcal{A}( B^-\to \overline D^0\pi^- \overline K^0)\nonumber\\&-\mathcal{A}( \overline B^0\to  D^-\pi^+\overline K^0)+\sqrt{2}\,\mathcal{A}( \overline B^0\to  \overline D^0\pi^0 \overline K^0)\nonumber\\&+\mathcal{A}( \overline B^0\to \overline D^0\pi^+ K^-)\big]=0,
\end{align}
\begin{align}
{ SumI_-}\,[\overline B^0,\overline D^0,\pi^+,\overline K^0]=&\mathcal{A}( \overline B^0\to  D^-\pi^+\overline K^0)-\sqrt{2}\,\mathcal{A}( \overline B^0\to  \overline D^0\pi^0 \overline K^0)\nonumber\\&-\mathcal{A}( \overline B^0\to \overline D^0\pi^+ K^-)=0,
\end{align}
\begin{align}
{ SumI_-}\,[B^-,D^-,\pi^+,\overline K^0]=&-\sqrt{2}\,\mathcal{A}( B^-\to D^-\pi^0 \overline K^0)-\mathcal{A}( B^-\to D^-\pi^+ K^-)\nonumber\\&+\mathcal{A}( \overline B^0\to  D^-\pi^+\overline K^0)=0,
\end{align}
\begin{align}
{ SumI_-}\,[B^-,\overline D^0,\pi^0,\overline K^0]=&\,\mathcal{A}( B^-\to D^-\pi^0 \overline K^0)-\mathcal{A}( B^-\to \overline D^0\pi^0 K^-)\nonumber\\&+\sqrt{2}\,\mathcal{A}( B^-\to \overline D^0\pi^- \overline K^0)+\mathcal{A}( \overline B^0\to  \overline D^0\pi^0 \overline K^0)=0,
\end{align}
\begin{align}
{ SumI_-}\,[B^-,\overline D^0,\pi^+, K^-]=&\,\mathcal{A}( B^-\to D^-\pi^+ K^-)-\sqrt{2}\,\mathcal{A}( B^-\to \overline D^0\pi^0 K^-)\nonumber\\&+\mathcal{A}( \overline B^0\to \overline D^0\pi^+ K^-)=0,
\end{align}
\begin{align}
{ SumI_-}\,[B^-,\overline D^0,\overline K^0,\eta_8]=&\,\mathcal{A}( B^-\to D^- \overline K^0\eta_8)-\mathcal{A}( B^-\to \overline D^0 K^-\eta_8)\nonumber\\&+\mathcal{A}( \overline B^0\to  \overline D^0 \overline K^0\eta_8)=0,
\end{align}
\begin{align}
{ SumI_-^2}\,[B^-, D^-_s,\pi^+,\pi^+]=&-2\,\big[-2\,\mathcal{A}( B^-\to D^-_s \pi^0\pi^0)+\mathcal{A}(B^-\to D^-_s \pi^-\pi^+)\nonumber\\&+\mathcal{A}( B^-\to D^-_s \pi^+\pi^-)+\sqrt{2}\,\mathcal{A}( \overline B^0\to D^-_s \pi^0\pi^+)\nonumber\\&+\sqrt{2}\,\mathcal{A}( \overline B^0\to D^-_s \pi^+\pi^0)\big]=0,
\end{align}
\begin{align}
{ SumI_-}\,[\overline B^0, D^-_s,\pi^+,\pi^+]=&-\sqrt{2}\,\mathcal{A}( \overline B^0\to D^-_s \pi^0\pi^+)-\sqrt{2}\,\mathcal{A}( \overline B^0\to D^-_s \pi^+\pi^0)=0,
\end{align}
\begin{align}
{ SumI_-}\,[B^-, D^-_s,\pi^+,\pi^0]=&-\sqrt{2}\,\mathcal{A}( B^-\to D^-_s \pi^0\pi^0)+\sqrt{2}\,\mathcal{A}(B^-\to D^-_s \pi^+\pi^-)\nonumber\\&+\mathcal{A}( \overline B^0\to D^-_s \pi^+\pi^0)=0,
\end{align}
\begin{align}
{ SumI_-}\,[B^-, D^-_s,\pi^+,\eta_8]=&-\sqrt{2}\,\mathcal{A}( B^-\to D^-_s \pi^0\eta_8)+\mathcal{A}( \overline B^0\to D^-_s \pi^+\eta_8)=0,
\end{align}
\begin{align}
{ SumI_-}\,[B^-, D^-_s,K^+,\overline K^0]=&\,\mathcal{A}( B^-\to D^-_s K^0\overline K^0)-\mathcal{A}(B^-\to D^-_s K^+K^-)\nonumber\\&+\mathcal{A}( \overline B^0\to D^-_s K^+\overline K^0)=0,
\end{align}
\begin{align}
{ SumI_-^2}\,[\overline B^0_s, \overline D^0,\pi^+,\pi^+]=&-2\,\big[\,\sqrt{2}\,\mathcal{A}( \overline B^0_s\to D^- \pi^0\pi^+)+\sqrt{2}\,\mathcal{A}(\overline B^0_s\to D^- \pi^+\pi^0)\nonumber\\&-2\,\mathcal{A}( \overline B^0_s\to \overline D^0 \pi^0\pi^0)+\mathcal{A}( \overline B^0_s\to \overline D^0 \pi^-\pi^+)\nonumber\\&+\mathcal{A}( \overline B^0_s\to \overline D^0 \pi^+\pi^-)\big]=0,
\end{align}
\begin{align}
{ SumI_-}\,[\overline B^0_s, D^-,\pi^+,\pi^+]=&-\sqrt{2}\,\mathcal{A}( \overline B^0_s\to D^- \pi^0\pi^+)-\sqrt{2}\,\mathcal{A}(\overline B^0_s\to D^- \pi^+\pi^0)=0,
\end{align}
\begin{align}
{ SumI_-}\,[\overline B^0_s, \overline D^0,\pi^+,\pi^0]=&\,\mathcal{A}( \overline B^0_s\to D^- \pi^+\pi^0)-\sqrt{2}\,\mathcal{A}( \overline B^0_s\to \overline D^0 \pi^0\pi^0)\nonumber\\&+\sqrt{2}\,\mathcal{A}( \overline B^0_s\to \overline D^0 \pi^+\pi^-)=0,
\end{align}
\begin{align}
{ SumI_-}\,[\overline B^0_s, \overline D^0,\pi^+,\eta_8]=&\,\mathcal{A}( \overline B^0_s\to D^- \pi^+\eta_8)-\sqrt{2}\,\mathcal{A}( \overline B^0_s\to \overline D^0 \pi^0\eta_8)=0,
\end{align}
\begin{align}
{ SumI_-}\,[\overline B^0_s, \overline D^0,K^+,\overline K^0]=&\,\mathcal{A}( \overline B^0_s\to D^- K^+\overline K^0)+\mathcal{A}( \overline B^0_s\to \overline D^0 K^0\overline K^0)\nonumber\\&-\mathcal{A}( \overline B^0_s\to \overline D^0 K^+K^-)=0,
\end{align}
\begin{align}
{ SumI_-}\,[\overline B^0_s,  D^-_s,\pi^+,K^+]=&-\sqrt{2}\,\mathcal{A}( \overline B^0_s\to D^-_s \pi^0 K^+)+\mathcal{A}( \overline B^0_s\to  D^-_s \pi^+ K^0)=0.
\end{align}

\end{appendix}

\end{document}